\documentclass[%
 aip,
 amsmath,amssymb,
 reprint,%
]{revtex4-1}

\usepackage{graphicx}
\usepackage{dcolumn}
\usepackage{bm}

\usepackage[utf8]{inputenc}
\usepackage[T1]{fontenc}
\usepackage{mathptmx}
\usepackage{xcolor}

\usepackage[normalem]{ulem}

\graphicspath{{figures/}}

\begin{document}

\preprint{AIP/123-QED}

\title[Causality indices for bivariate time series data]{Causality indices for bivariate time series data:\\a comparative review of performance}

\author{Tom Edinburgh}
\email{te269@cam.ac.uk}
\author{Stephen J. Eglen}
\affiliation{Department of Applied Mathematics and Theoretical Physics, University of Cambridge, Cambridge CB3 0WA, UK}
\author{Ari Ercole}
\affiliation{Cambridge Centre for Artificial Intelligence in Medicine and Division of Anaesthesia, Department of Medicine, University of Cambridge, Cambridge CB2 0QQ, UK}

\begin{abstract}
Inferring nonlinear and asymmetric causal relationships between multivariate longitudinal data is a challenging task with wide-ranging application areas including clinical medicine, mathematical biology, economics and environmental research. A number of methods for inferring causal relationships within complex dynamic and stochastic systems have been proposed but there is not a unified consistent definition of causality in the context of time series data. We evaluate the performance of ten prominent causality indices for bivariate time series, across four simulated model systems that have different coupling schemes and characteristics. Pairwise correlations between different methods, averaged across all simulations, show there is generally strong agreement between methods, with minimum, median and maximum Pearson correlations between any pair (excluding two similarity indices) of 0.298, 0.719 and 0.955 respectively. In further experiments, we show that these methods are not always be invariant to real-world relevant transformations (data availability, standardisation and scaling, rounding error, missing data and noisy data). We recommend transfer entropy and nonlinear Granger causality as particularly strong approaches for estimating bivariate causal relationships in real-world applications. Both successfully identify causal relationships and a lack thereof across multiple simulations, whilst remaining robust to rounding error, at least 20\% missing data and small variance Gaussian noise. Finally, we provide flexible open-access Python code for computation of these methods and for the model simulations. 
\\[0.4em]
Author-accepted manuscript, accepted for publication in AIP Chaos on 01/06/2021.
\end{abstract}

\maketitle

\begin{quotation}
Quantifying causal relationships between longitudinal observations of a complex system is essential to an understanding of the interactions between sub-components of the system and is subsequently key to building better and more parsimonious models \cite{Granger1969-bj, Runge2018-dk}. In many real-world applications, we are rarely able to access or describe an underlying graphical network of these interactions \textit{a priori}, and we are typically limited to observing simultaneously recorded variables from each subsystem as a multivariate time series. Two key properties that are widely regarded as crucial in defining causal relationships are: that the effect is temporally preceded by the cause, and that external changes to values of the causal variable propagate to values of the effect variable and do not break the causal structure \cite{Eichler2012-tw}. Correlation or synchronisation in these multivariate time series does not necessarily imply a causal relationship between variables, and counter-examples are easy to find \cite{Aldrich1995-iz}. Further, a lack of correlation does not imply a lack of causality, and a reliance on correlation-based measures may result in nonlinear causal relationships being obscured, e.g. Ref~\onlinecite{Sugihara2012-lv}. In recent decades, various mathematical frameworks \cite{Granger1969-bj, Sims1972-gj, Schreiber2000-hz} have been described to allow identification of nonlinear (and asymmetric) causal structure within complex systems, primarily driven by domain-specific applications, from diverse application areas including as statistical economics \cite{Granger1969-pv, Geweke1984-er}, climate science \cite{Zhang2011-fc, Runge2019-zq, Runge2019-nm} and computational neuroscience \cite{Gray1989-zx, Seth2015-rp}.
\end{quotation}

\section{\label{sec:introduction}Introduction}

\begin{figure}[t]
  \centering
  \includegraphics[width=8.5cm]{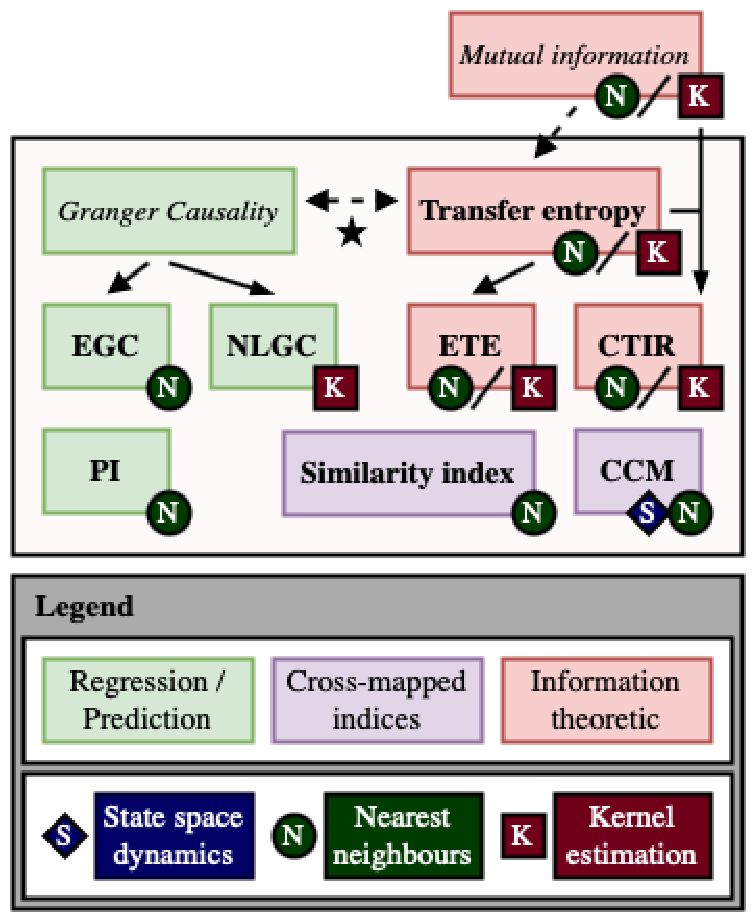}
  \caption{Causality indices described in this paper, which represent a widely-used but non-exhaustive subset of this field of research. The indices are as follows (where GC is Granger causality): extended GC (EGC), nonlinear GC (NLGC), predictability improvement (PI), transfer entropy (TE), effective transfer entropy (ETE), coarse-grained transinformation rate (CTIR), similarity indices (SI), convergent cross mapping (CCM). We classify these indices into three categories, and highlight commonalities between the approaches and their estimation (state space dynamics, nearest neighbour computation, kernel estimation).}
  \label{fig:classification}
\end{figure}

No general method exists to identify causal structure within complex systems, and there is no single consistent and unifying notion of quantitative causality estimation for time series data. Published methods can be broadly categorised into the following groups:
\begin{enumerate}
    \item regression-based indices that use `recent history' vectors as predictors in a model (e.g. Granger causality),
    \item information-theoretic indices that build upon ideas of conditional mutual information (e.g. transfer entropy),
    \item indices based on state space dynamics, such as local neighbourhoods and trajectories (e.g. convergent cross mapping), 
    \item graphical models that scale causal inference estimation to high-dimensional multivariate systems for causal identification.
\end{enumerate}There exist common themes between these methods, and membership within these groups is sometimes somewhat blurred. Figure~\ref{fig:classification} sets out key properties and similarities between methods from groups 1-3. Previous reviews of the literature \cite{Hlavackova-Schindler2007-lm, Eichler2013-pb, Papana2013-ul, Palachy2019-kx} typically focus on a subset of methods from one of these groups. The suitability, interchangeability and performance of published methods, particularly where they span more than one of these groups, has received relatively little attention. In this work, we identified and assessed a widely used subset of indices for directed bivariate causality inference, concentrating on methods involving univariate embeddings to describe the recent history of the system (Section \textbf{II.}). A review of such methods has been published previously by Lungarella \textit{et al.} \cite{Lungarella2007-vs}. We reproduce these results for the original and newer methods. We also extend this work by proposing a set of modifications that can be made to simulated data prior to causal estimation, in order to investigate sensitivity of each method to data availability, scaling, missing data, rounding and Gaussian noise (Section \textbf{III.}). Each of these reproduces phenomena that often occur in real-world data, such as when instruments have a fixed measurement precision and data is reported with rounding error. We believe these tests should provide in-depth benchmarking criteria for new proposed methodologies. Finally, we summarise the strengths and weaknesses of these approaches, and identify key areas of further research (Section \textbf{IV.}).

\section{\label{sec:methods}Methods}

We observe a complex system as a set of variables within a multivariate time series. The time series $X = \left(x_1, \ldots, x_T\right)$ and $Y = \left(y_1, \ldots, y_T\right)$ describe a bivariate system with state $\mathbf{s}_t = \left(x_t,y_t\right)$ at time $t$. A critical implicit assumption is that the series has unit time, or equivalently that the data is observed at a fixed constant frequency. Underpinning all methods is the key assumption that cause directly precedes effect. As a result, a preliminary step is the construction of time-delay embedding vectors $\mathbf{x}_{t}^{m, \tau}$ in $m$-dimensional state space $\mathcal{X} \cong \mathbb{R}^m$. An important distinction in the defined methods is whether their theoretical basis is stochastic or deterministic. Both employ the same time-delay vectors, though only in deterministic systems are the vectors $\mathbf{x}_{t}^{m, \tau}$ considered elements within an $m$-dimensional state space $\mathcal{X} \cong \mathbb{R}^m$. We construct equivalent embedding vectors $\mathbf{y}_{t}$ (and state space $\mathcal{Y} \cong \mathbb{R}^m$) for $Y$, and a joint embedding vectors $\mathbf{z}_{t}$ (and state space $\mathcal{Z} \cong \mathbb{R}^{2m}$):
\begin{eqnarray*}
    \mathbf{x}_{t} = \mathbf{x}_{t}^{m, \tau} = \left( x_{t - (m - 1)\tau}, x_{t - (m - 2)\tau}, \ldots, x_{t - \tau}, x_t \right)^\prime \in \mathcal{X}\\
    \mathbf{z}_{t} = \begin{pmatrix}\mathbf{x}_{t}\\ \mathbf{y}_{t} \end{pmatrix} \in \mathcal{Z},~~~~~t = (m - 1)\tau + 1,\ldots,T
    \label{eq:embedding}
\end{eqnarray*}

In practice, many real-world systems are stochastic, with some level of noise or randomness in at least part of the system. A further assumption for stochastic causality estimation is that of separability, which states that there is unique information about the effect variable that is contained only within the causal variable. The standard approach here is to describe or model the current value $x_{t}$ of $X$ as conditional upon upon the `recent history' joint embedding vector $\mathbf{z}_{t-1}$ (full model). Separability means that removing the causal variable $Y$ eliminates the information it contains about the effect $X$, which we observe either by identifying non-zero coefficients in the full model or constructing a reduced model, conditioned only upon $\mathbf{x}_{t-1}$. These methods are generally described with time index shifted $t \mapsto t+1$, though the interpretation (`current' and `recent history') remains the same.

Granger causality \cite{Granger1969-bj} (GC), a notable and popular method for causality estimation in time series, fits autoregressive models on the time series to this end. Extensions of GC to nonlinear systems include a locally linear version called \textbf{extended Granger causality} \cite{Chen2004-vu} (EGC) and \textbf{nonlinear Granger causality} \cite{Ancona2004-cr} (NLGC), which performs a `global' nonlinear autoregression using radial basis functions (RBFs). \textbf{Predictability improvement} \cite{Feldmann2004-yt} (PI) is another locally constant linear regression of `recent history' embeddings, which measures a reduction in mean squared error when $\mathbf{z}_t$ is used for predicting a `horizon' value $x_{t+h}$ instead of $\mathbf{x}_t$ alone. 

Information theory is a natural framework for describing causal relationships. \textbf{Transfer entropy} \cite{Schreiber2000-hz} (TE) measures deviation from the generalised Markov property $p\left(x_{t+1}|\mathbf{x_t}\right) = p\left(x_{t+1}|\mathbf{x_t}, \mathbf{y_t}\right)$ as a conditional mutual information. With weak coupling and limited data, transfer entropy can suffer from finite sample effects and \textbf{effective transfer entropy} \cite{Marschinski2002-zh} (ETE) corrects for this using shuffled versions of the causal variable. TE reduces to vanilla GC under the assumption of Gaussian variables \cite{Barnett2009-se} ($\star$, Figure~\ref{fig:classification}), and non-zero GC implies violation of the generalised Markov property and non-zero TE \cite{Marinazzo2008-os}. \textbf{Coarse-grained transinformation rate} \cite{Palus2001-po} (CTIR) is based upon `coarse-grained entropy rates', and measures the rate of net information flow, averaged over different lags $\tau$. Often the difficulty in information theoretic methods (described in depth in Ref~\onlinecite{Hlavackova-Schindler2007-lm}) is the robust estimation of joint probabilities or entropy values, which in turn form building blocks for these methods. We use a histogram binning partition (H) and the (hypercube) Kraskov-St{\"o}gbauer-Grassberger (KSG) estimate \cite{Kraskov2004-sf}, which is a technique involving $k$-nearest neighbour statistics. All information theoretic computation here is in `nats' (logarithm base $e$).

Fully deterministic dynamical systems, which evolve according to a differential equation or difference equation, do not necessarily satisfy the separability condition. In these systems, $x_t$ can often be reformulated as a function of only past values of $X$, which makes the potential causal role of $Y$ in the coupled system less clear, as highlighted by Granger \cite{Granger1969-bj}. Causal relationships in a coupled deterministic system are instead observed via the event that each variable belongs to a shared attractor manifold $A \subset \mathcal{Z}$. A consequence of Takens' embedding theorem \cite{Takens1981-tt} is that the `library of historical behaviour' of $X$ preserves the topology of $A$ and, by transitivity, local neighbourhoods in $\mathcal{X}$ those in $\mathcal{Y}$ and vice versa \cite{Sugihara2012-lv}. It is possible to detect unidirectional causal influence, where only the dynamics of a causal variable propagate to the response variable in this way. Sugihara \textit{et al.} \cite{Sugihara2012-lv} argue that the inferred direction of unidirectional causal influence is counter-intuitively reversed (i.e. cross mapping from $\mathcal{X}$ to $\mathcal{Y}$ reveals causal influence from $Y$ to $X$).

The key assumption of cross mapped indices is that causal relationships are observed in the similarity between sets of (subscript) indices denoting the nearest neighbours for each set of embedding vectors, which can be mapped from one variable to the other to reveal interdependency. This is the idea behind the \textbf{similarity indices}: two similarity indices we test here, denoted $\textnormal{SI}_{Y\rightarrow X}^{(1)}$ and $\textnormal{SI}_{Y\rightarrow X}^{(2)}$, are $H(\mathcal{X}|\mathcal{Y})$ in Ref~\onlinecite{Arnhold1999-si} and $E(\mathcal{X}|\mathcal{Y})$ in Ref~\onlinecite{Bhattacharya2003-cz} respectively. \textbf{Convergent cross mapping} \cite{Sugihara2012-lv} (CCM) computes the correlation $\rho$ between the cross mapped estimate and the true value, with convergence in $\rho$ as $T$ increases ``a key property that distinguishes causation from simple correlation'' \cite{Sugihara2012-lv}. 

\begin{table*}
\caption{Causality indices in this review and their parameters. The indices are as follows (where GC is Granger causality): extended GC (EGC), nonlinear GC (NLGC), predictability improvement (PI), transfer entropy (TE), effective transfer entropy (ETE), coarse-grained transinformation rate (CTIR), similarity indices (SI) and convergent cross mapping (CCM). Table~S.I (\textbf{Supplementary materials}) provides more detail on the parameter choices for individual simulation results.}
\begin{tabular}{l|l|l|l|l|l}
    \noalign{\hrule height 2pt}
    & Method & \multicolumn{2}{l|}{Parameters / other choices} & Notes / suggestions & Values used here \\
    \noalign{\hrule height 1.2pt}
    & All & $T$ & Time series length & Depends on data availability & $10^p,~p = 3,4,5$ \\
    Embedding &  & $h$ & Time horizon value & Normally $h = 1$, generalised to $h \geq 1$ (in PI) & $h = 1$ \\
    & All $\backslash$ CTIR & $m$ & Embedding dimension & `Optimal' \cite{Kennel1992-ej} vs `empirical' ($m = 1,\mathinner{{\ldotp}{\ldotp}{\ldotp}},5$) & $m = 1$ or $2$ \\ 
    &  & $\tau$ & Time-delay lag & `Optimal' \cite{Fraser1986-bc} vs `empirical' ($\tau = 1,2,3$) & $\tau = 1$ \\ 
    \hline
    & \textbf{EGC} \cite{Chen2004-vu} & \multicolumn{2}{l|}{Nearest neighbour metric} & $\ell_p$, may depend on state space / distribution & $\ell_1$ \\ 
    &  & $L$ & No. of neighbourhoods & $L = 100$ in Refs. \onlinecite{Lungarella2007-vs, Chen2004-vu} (depends on $T$) & $L = 20$ or $100$ \\
    &  & $\delta$ & Neighbourhood size & Compute EGC for $\delta \downarrow 0$ (Ref \onlinecite{Chen2004-vu}) & Various (Table~S.I)  \\
    & \textbf{NLGC} \cite{Ancona2004-cr} & \multicolumn{2}{l|}{Radial basis function (RBF)} & Gaussian RBFs in Refs. \onlinecite{Lungarella2007-vs, Ancona2004-cr} & Gaussian \\
    Regression &  & $P$ & No. of RBFs & e.g. gap statistics \cite{Tibshirani2001-tp} & Various (Table~S.I) \\
    error &  & $\mathbf{x}_{\rho}$ & Gaussian RBF centers & via $k$-means or fuzzy $c$-means clustering & via $k$-means \\
    &  & $\sigma^2$ & Gaussian RBF variance & \textit{A priori} fixed e.g. $\sigma^2 = 0.05$ in Refs \onlinecite{Lungarella2007-vs, Ancona2004-cr} & $\sigma^2 = 0.05$ \\  
    & \textbf{PI} \cite{Feldmann2004-yt} & \multicolumn{2}{l|}{Nearest neighbour (NN) metric} & $\ell_p$, may depend on state space / distribution & $\ell_2$ \\
    &  & $R$ & No. of NNs & \textit{A priori} unclear, e.g. $R = 1, 10$ in Refs. \onlinecite{Lungarella2007-vs, Feldmann2004-yt} & $R = 1$ or $10$ \\
    &  & $h$ & Time horizon value & As above, e.g. $h = 1$ in Refs. \onlinecite{Lungarella2007-vs, Feldmann2004-yt} & $h = 1$ \\
    \hline
    Information & Estimation & \multicolumn{2}{l|}{Estimation method} & e.g. KSG, histogram partition & Both (H / KSG) \\
    theory &  & \multicolumn{2}{l|}{Nearest neighbour metric (KSG)} & $\ell_\infty$ (for hypercube dimensions) \cite{Kraskov2004-sf} & $\ell_\infty$ \\
    &  & $k$ & No. of NNs (KSG) & Small values e.g. $k = 2,3,4$ \cite{Kraskov2004-sf} & $k = 4$ \\
    &  & $N$ & No. of bins (histogram) & e.g. via minimum description length \cite{Rissanen1978-jp, Hall1988-mg} & $N = 8$ \\
    & \textbf{TE} \cite{Schreiber2000-hz} & n/a & & No parameters besides estimation (above) & n/a \\
    & \textbf{ETE} \cite{Marschinski2002-zh} & $N_{\textnormal{shuffle}}$ & No. of shuffled $X$ or $Y$ & \textit{A priori} unclear, single shuffle in Ref \onlinecite{Marschinski2002-zh} & $N_{\textnormal{shuffle}} = 10$ \\
    & \textbf{CTIR} \cite{Palus2001-po} & $\tau_{\max}$ & Max time-delay lag & $\tau_{\max}: I(x_t,x_{t+\tau'}) \approx 0,~ \forall\tau'\geq\tau_{\max}$ \cite{Palus2001-po} & $\tau_{\max} = 5$ or $20$ \\  
    & & $\tau_I,~\epsilon_I$ & For estimation of $\tau_{\max}$ & $\tau_{\max} = \min_{\tau'}\{\tau' \leq \tau_I : I(x_t,x_{t+\tau'}) < \epsilon_I\}$ & Unused, fixed $\tau_{\max}$ \\  
    \hline
    Cross & \textbf{SI} \cite{Arnhold1999-si, Bhattacharya2003-cz} & \multicolumn{2}{l|}{Nearest neighbour (NN) metric} & $\ell_p$, may depend on state space / distribution & $\ell_2$ \\
    mapped &  & $R$ & No. of NNs & \textit{A priori} unclear, e.g. $R = 10$ in Refs. \onlinecite{Arnhold1999-si,Bhattacharya2003-cz} & Various (Table~S.I) \\
    & \textbf{CCM} \cite{Sugihara2012-lv} & \multicolumn{2}{l|}{Nearest neighbour metric} & $\ell_p$, may depend on state space / distribution & $\ell_2$ \\ 
    &  & $T_{\max}$ & Max. segment length & Convergence: compute $\rho$ for $T \uparrow$ $T_{\max}$ \cite{Sugihara2012-lv} & $T_{\max} = T$ \\
    &  & $n_T$ & No. segments of size $T$ & $\rho$ values averaged across $n_T$ segments, size $T$ & $n_T = 40$ \\
    &  & $\rho_\infty$ & Converged CCM value & 
    $\rho_{T_{\max}}$ in Ref \onlinecite{Clark2015-nv} or fit exponential regression \cite{Monster2016-wy} & $\rho_{T_{\max}}$ (if $\downarrow$ holds) \\
    &  & $\delta_\rho$ & Convergence tolerance & `Converged' if $\rho_\infty - \rho_{m + 2} > \delta_\rho$ & $\delta_\rho = 0.05$ \\
    \noalign{\hrule height 2pt}
\end{tabular}
\label{tab:parameters}
\end{table*}

\section{\label{sec:results}Results}

Our results are split into two parts. First, we reproduce the results from Ref~\onlinecite{Lungarella2007-vs}, evaluating the performance of all methods including the additional CTIR and CCM, plus ETE using histogram binning and TE using KSG. In these simulations, we choose the same simulation model parameters and causality index parameters as in Ref~\onlinecite{Lungarella2007-vs} (Table~\ref{tab:parameters}). In the second part, we investigate sensitivity to common issues relevant to real-world data, using the Ulam lattice system to illustrate these.

\subsection{\label{sec:nsim2}Numerical simulations}

\begin{table*}
\caption{Brief summary of the characteristics of each numerical simulation model system and parameters (equations \ref{eq:sim1}-\ref{eq:sim4}). The difference between identical and non-identical H{\' e}non bidirectional maps is the value of $b_y$ ($b_y=b_x$ for identical maps and $b_y<b_x$ for non-identical maps). In each simulation, the first $10^5$ iterations were discarded as transients ($10^4$ for linear process). Each simulation is initialised randomly but seeded for reproducibility. The coupling parameters were incremented by $0.01$ in all cases, for each of 10 independent runs and all indices. We use the following abbreviations in this table: I - identical maps; NI - non-identical maps; L \& S - linear and stochastic; NL \& D \& C - non-linear, deterministic and chaotic.}
\begin{tabular}{ l|l|l|l|l|l }
    \noalign{\hrule height 2pt}
    Simulation & Coupling & Dynamics & $T = 10^p$ & Simulation model parameters & Coupling strength \\
    \noalign{\hrule height 1.2pt}
    Linear process & $X\leftarrow Y$ & L \& S & $p = 4$ & $b_x = 0.8,~b_y = 0.4,~\sigma_x^2 = \sigma_y^2 = 0.2$ & $\lambda \in [0,1]$ \\ 
    Ulam lattice & $X\rightarrow Y$ & NL \& D \& C & $p = 3, 5$ & $N_L = 100$ (size of lattice) & $\lambda \in [0,1]$ \\ 
    H{\' e}non uni-d & $X\rightarrow Y$ & NL \& D \& C & $p = 3, 4, 5$ & $a = 1.4,~b_x = 0.3,~b_y = 0.3$ & $\lambda \in [0,1]$ \\ 
    H{\' e}non bi-d (I, NI) & $X\leftrightarrow Y$ & NL \& D \& C & $p = 4$ & $a = 1.4,~b_x = 0.3,~b_y = 0.3$ or $0.1$ & $\lambda_{xy},\lambda_{yx} \in [0,0.4]$ \\ 
    \noalign{\hrule height 2pt}
\end{tabular}
\label{tab:simulations}
\end{table*}

We investigate performance on four simulated model systems (Table~\ref{tab:simulations}). In each simulation, we assess the causality estimates of each method by varying the coupling strength $\lambda$. These simulated systems are widely studied in chaos theory, e.g. Ref \onlinecite{Henon1976-kw}, and also appear elsewhere in the literature, e.g. Ulam lattice in Ref \onlinecite{Schreiber2000-hz}.

\begin{figure*}[t]
  \centering
  \includegraphics[width=17cm]{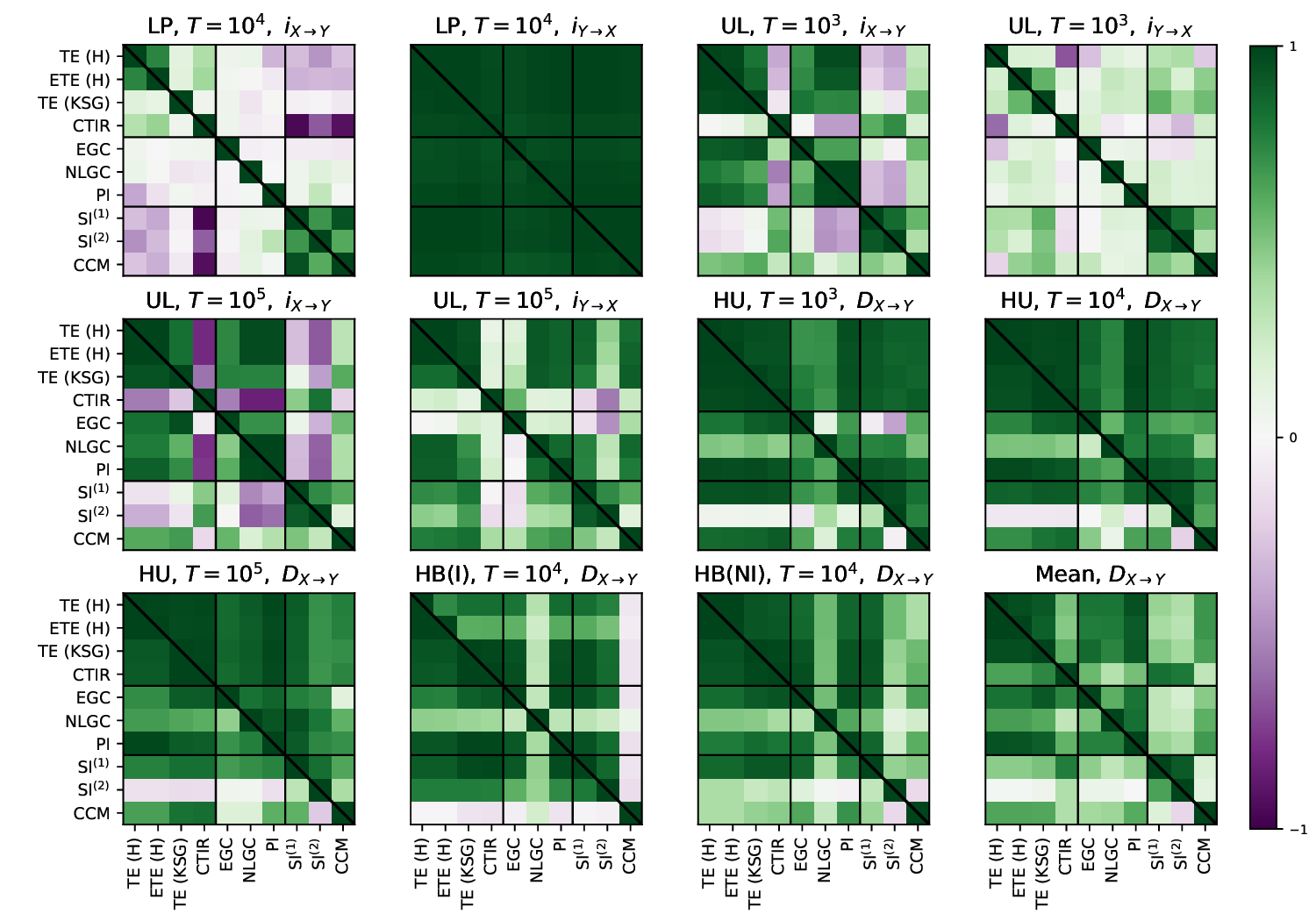}
  \caption{Correlations between each of the causality indices, for all simulations: linear process (LP), Ulam lattice (UL) and H{\'e}non unidirectional maps (HU), identical/non-identical H{\'e}non bidirectional maps maps (HB (I)/HB (NI)). For several of the simulated systems, simulations were repeated for increasing data size $T$. In each subplot, the lower left half below the diagonal shows the Pearson correlation between each pair of indices (across all runs and values of $\lambda$) and the upper right half shows the rank-based Spearman correlation. Both $i_{X\rightarrow Y}$ and $i_{Y\rightarrow X}$ are shown for LP and UL simulations but only $D_{X\rightarrow Y} = i_{X\rightarrow Y}- i_{Y\rightarrow X}$ was computed for HU and HB. In the final bottom right subplot, we average correlations in $D_{X\rightarrow Y}$ for each simulation, weighting each simulation equally.}
  \label{fig:corr}
\end{figure*}

\begin{figure*}[t]
  \centering
  \includegraphics[width=17cm]{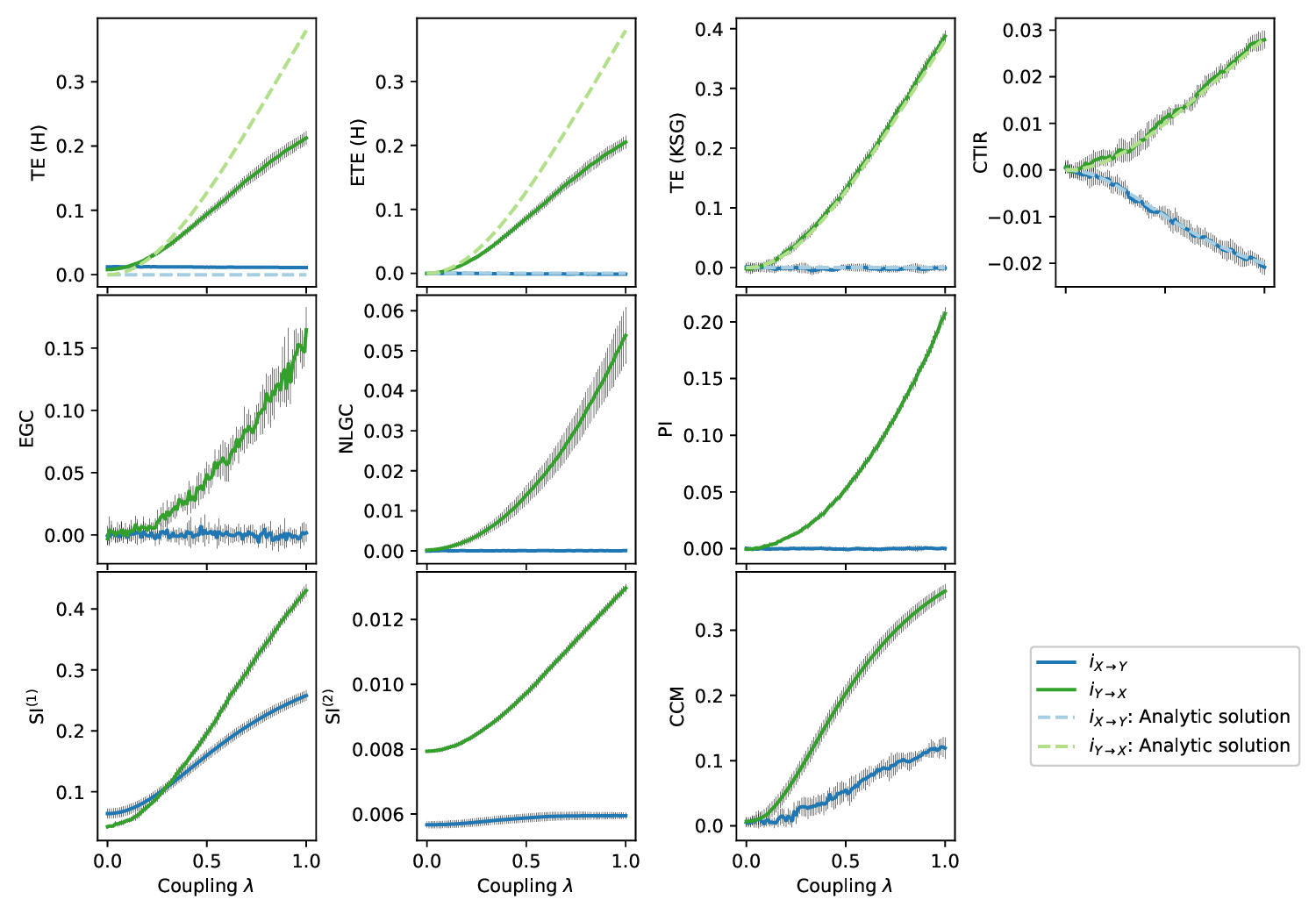}
  \caption{Linear Gaussian processes with $T = 10^4$ data points and unidirectional ($Y\!\rightarrow\! X$) coupling. Error bars report $\pm$1 empirical standard deviation from mean values, after 10 independent simulations from the LP system. Simulation parameters are given in Table~\ref{tab:simulations} and parameters for each causality index are given in Table~S.I. We have derived analytic solutions for all the information theoretic indices: TE (H), ETE (H), TE (KSG) and CTIR, which are shown as dashed lines. Note that for TE (KSG) and CTIR, the computational results match the analytic solutions almost exactly, and the dashed lines overlay the solid.}
  \label{fig:lp}
\end{figure*}

\begin{figure*}[p]
  \centering
  \includegraphics[width=16cm]{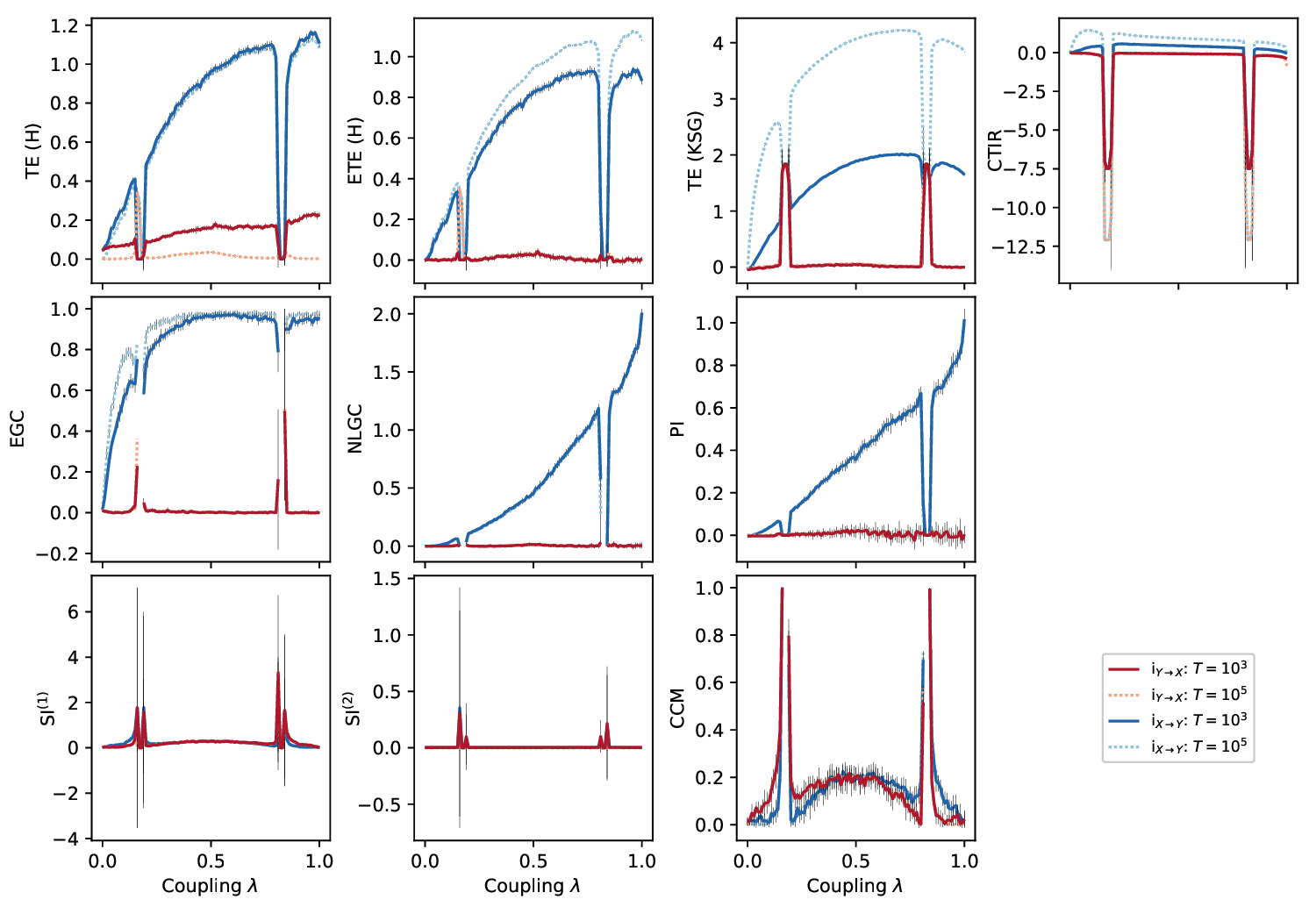}
  \includegraphics[width=16cm]{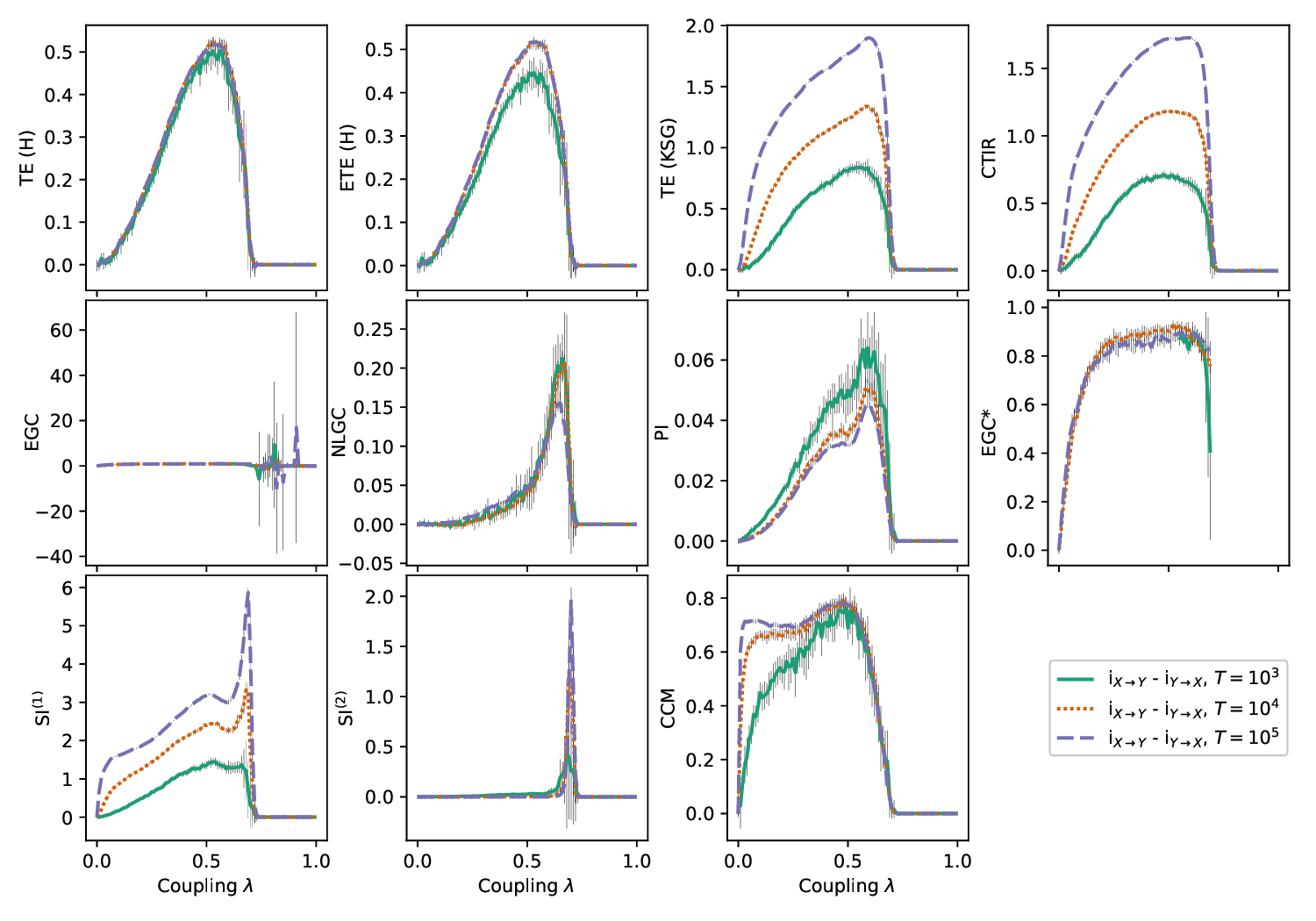}
  \caption{Ulam lattice (top half) and H{\' e}non unidirectional map (bottom half) simulation results, both with ($X\!\rightarrow\! Y$) coupling, with varying $T$. For the latter, we show the net directed index $D_{X\rightarrow Y} = i_{X\rightarrow Y} - i_{Y\rightarrow X}$. Error bars report $\pm$1 standard deviation from mean values, after 10 independent simulations of each map. Simulation parameters are given in Table~\ref{tab:simulations} and parameters for each causality index are given in Table~S.I. Due to extreme results in the EGC index when the system exhibits synchrony ($\lambda > 0.7$ for HU), we set these values set to NA and repeat the plot (EGC$*$).}
  \label{fig:hu}
\end{figure*}

\begin{figure*}[p]
  \centering
  \includegraphics[width=17cm]{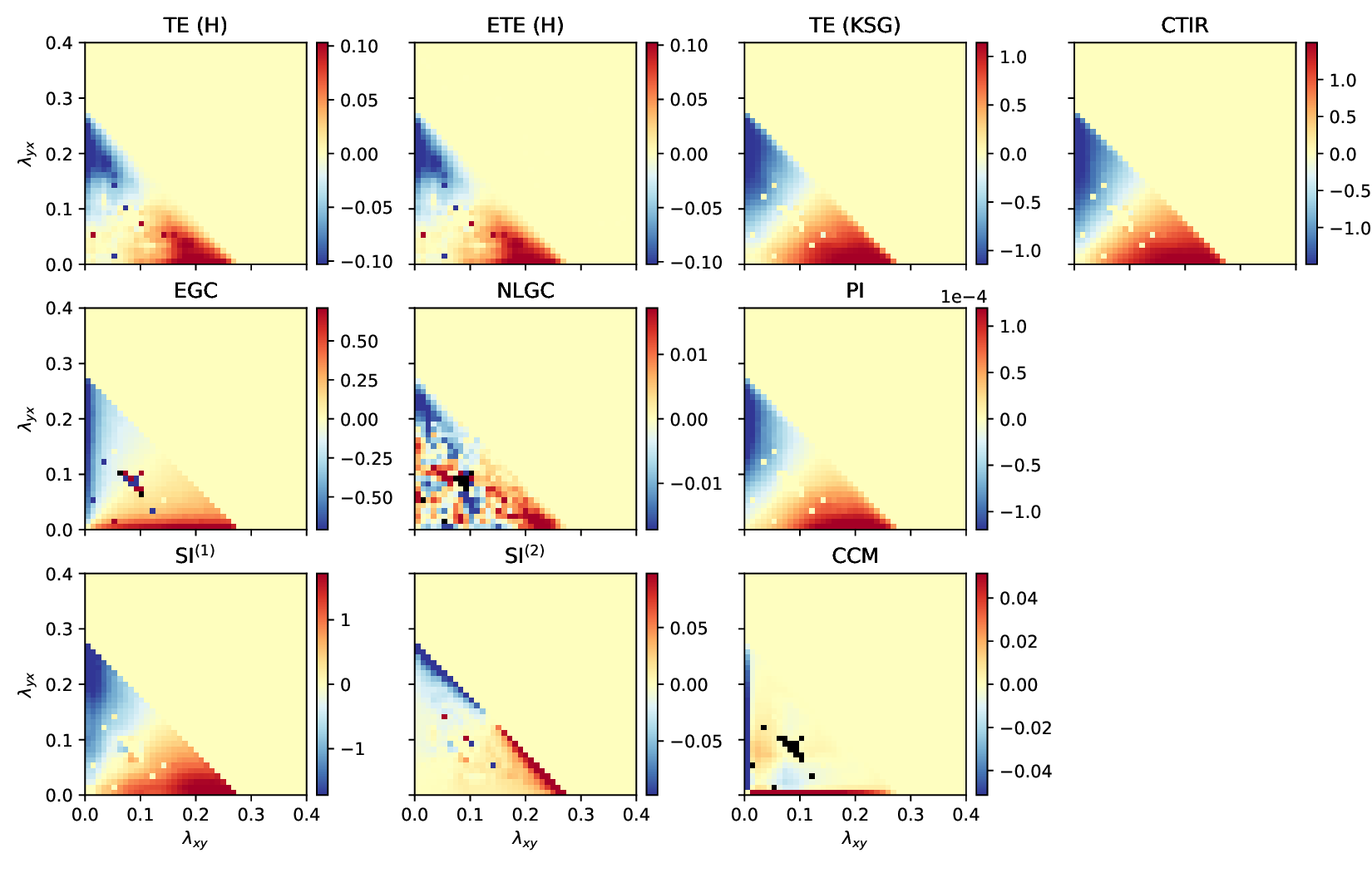}
  \includegraphics[width=17cm]{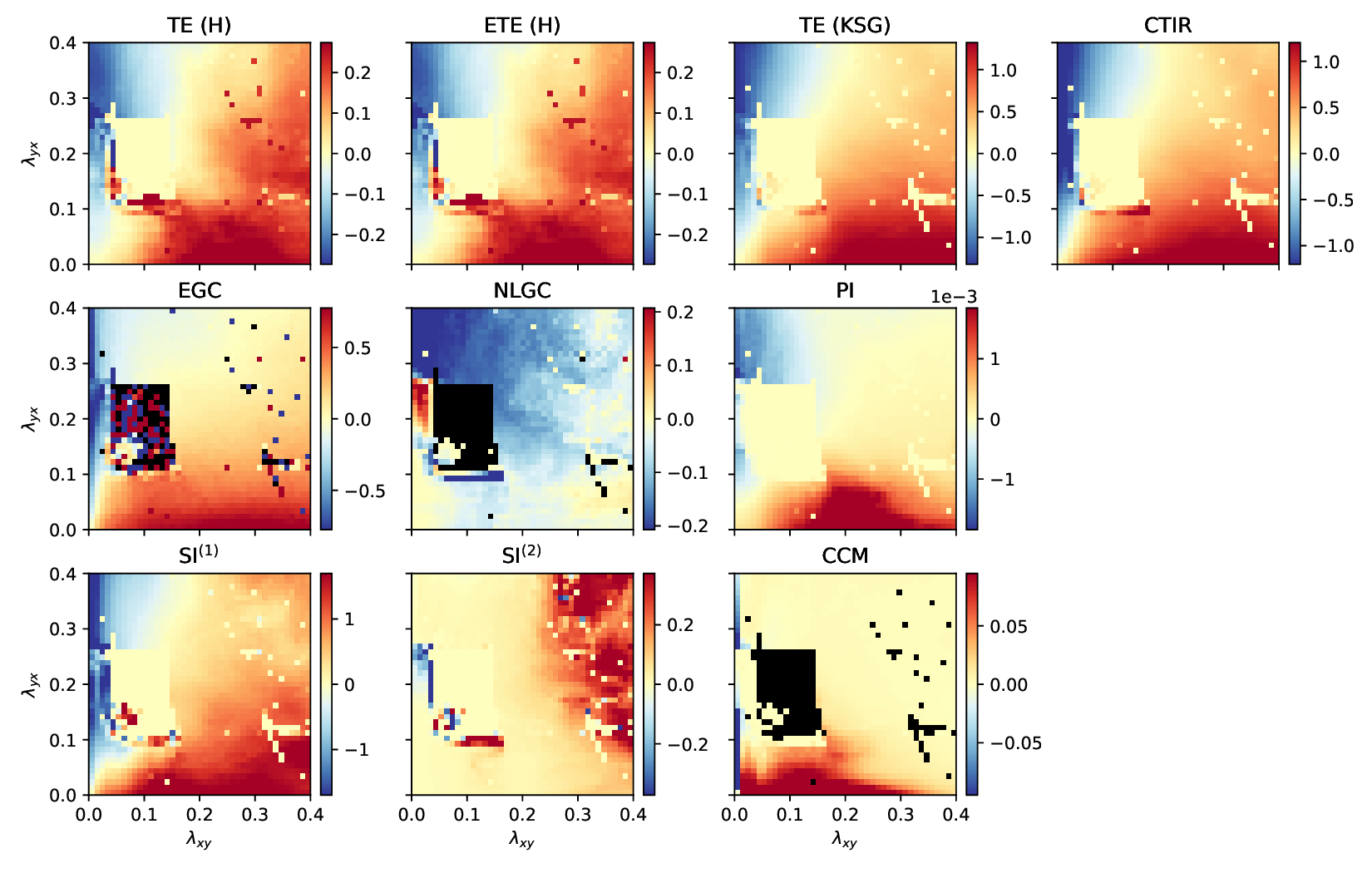}
  \caption{H{\' e}non map simulation results, with $T = 10^4$~data points and bidirectional coupling, for both identical maps (top half) and non-identical maps (bottom half). We show the net directed index $D_{X\rightarrow Y} = i_{X\rightarrow Y} - i_{Y\rightarrow X}$, averaged across 10 independent simulations. Colorbar limits are capped by percentiles, so as to minimise the effect of extreme values in data visualisation. We set these limits as $\pm \max (|p_1|, |p_{99}|)$ for identical maps and $\pm \max (|p_5|, |p_{95}|)$ for non-identical maps, where $p_i$ is the $i$\textsuperscript{th} percentile of results for that index. Simulation parameters are given in Table~\ref{tab:simulations} and parameters for each causality index are given in Table~S.I.}
  \label{fig:hb0}
\end{figure*}

\textbf{Linear process}:
\begin{eqnarray}
    x_{t+1} = b_x x_t + \lambda y_t + \epsilon_{x,t},~~~y_{t+1} = b_y y_t + \epsilon_{y,t} \label{eq:sim1} \\
    \epsilon_{x,t} \sim N(0,\sigma^2_x),~~~\epsilon_{y,t} \sim N(0,\sigma^2_y) \nonumber
\end{eqnarray}
\textbf{Ulam lattice}:
\begin{eqnarray}
    s_{t+1,l+1} = f(\lambda s_{t,l} + (1-\lambda)s_{t,l+1}),~~~l = 1,\ldots,N_L - 1 \\
    s_{t+1,1} = f(\lambda s_{t,N_L} + (1-\lambda)s_{t,1}),~~~f(s) = 2 - s^2 \nonumber \\
    x_t = s_{t,1},~~~y_t = s_{t,2} \nonumber
\end{eqnarray}
\textbf{H{\' e}non unidirectional map}:
\begin{eqnarray}
    x_{t+2} = a - x_{t+1}^2 + b_x x_t \\
    y_{t+2} = a - (\lambda x_{t+1} + (1-\lambda)y_{t+1})y_{t+1} + b_y y_t \nonumber 
\end{eqnarray}
\textbf{H{\' e}non bidirectional map}:
\begin{eqnarray}
    x_{t+2} = a - x_{t+1}^2 + \lambda_{yx}(x_{t+1}^2 - y_{t+1}^2) + b_x x_t \label{eq:sim4} \\
    y_{t+2} = a - y_{t+1}^2 + \lambda_{xy}(y_{t+1}^2 - x_{t+1}^2) + b_y y_t \nonumber
\end{eqnarray}
We reproduce figures for all simulations and methods in Figures~\ref{fig:lp}-\ref{fig:hb0}, and summarise our results in Figure~\ref{fig:corr}, which shows correlations between each pair of indices. For linear process and Ulam lattice simulations, we report causality estimates in both directions, i.e. $i_{X\rightarrow Y}$ and $i_{Y\rightarrow X}$ (where $i$ denotes any of the causality indices). For H{\' e}non maps, we instead use the directed index $D_{X\rightarrow Y} = i_{X\rightarrow Y} - i_{Y\rightarrow X}$, following Ref~\onlinecite{Lungarella2007-vs}. In general, all measures exhibit a small standard deviation relative to the absolute value of the index, indicating that random initial conditions during data simulation has at most a very small influence on the causal structure, when initial transients are discarded. Though we are able to replicate the findings in Ref~\onlinecite{Lungarella2007-vs} in most cases, we occasionally find minor differences between their results and ours. In particular, we sometimes find results of a similar profile but different magnitude, as $\lambda$ varies. We observe this for: EGC and linear processes; PI and all simulations, SI\textsuperscript{(1)} and Ulam lattice; TE and H{\' e}non unidirectional maps. Though we handle numerical outliers differently in our visualisation of results for H{\' e}non bidirectional maps, our results for these simulations appear largely comparable in magnitude and profile. There is no mention of a data standardisation step in Ref~\onlinecite{Lungarella2007-vs} and the results we report do not involve any pre-processing, though this did not appear to rectify these differences. In one notable difference between identical H{\' e}non bidirectional map results, Lungeralla \textit{et al.} \cite{Lungarella2007-vs} find a region in $\lambda$-space (namely $\{(\lambda_{xy},\lambda_{yx}):~\lambda_{xy} > 0.1,~\lambda_{yx} > 0.1,~\lambda_{xy}+\lambda_{yx} < 0.35\}$) in which they identify general synchronisation between $X$ and $Y$ and have difficulty estimating indices due to numerical instabilities, yet we do not observe this. We have followed the implementation in Ref~\onlinecite{Lungarella2007-vs} as closely as possible and it is unclear why these differences exist.

We knowingly deviated from the implementations in Ref~\onlinecite{Lungarella2007-vs} only in the case of NLGC, in which we preferred to use $k$-means rather than fuzzy $c$-means clustering to determine RBF centers, after finding similar or improved results but with a much reduced computational cost. Lungarella \textit{et al.} \cite{Lungarella2007-vs} note that NLGC is numerically unstable for `small' $T$ and computationally expensive for `large' $T$, which we suggest may be partly due to their use of fuzzy $c$-means. Little detail is provided about their implementation of this but it may perhaps be that an early stopping criteria sometimes forces a `poor  quality' clustering. Further, we found that the performance of NLGC in  H{\' e}non bidirectional map simulations improved significantly with a different set of NLGC parameters (e.g. $P = 50$ instead of $P = 10$), though we do not present these alternate results.

For the \textbf{linear process} (LP), the simplest simulation model, all indices show very strong positive correlation in the $Y\rightarrow X$ direction (Figure~\ref{fig:lp}). In the reverse $X\rightarrow Y$ direction, TE and CTIR both decrease with increasing $\lambda$, the cross mapped indices all show a marked increase and the remaining indices are approximately zero for all $\lambda$. This gives rise to patterns of positive and negative correlation between pairs of methods. As each $x_t$ or $y_t$ is a sum of Gaussian variables, we can derive theoretical values for Shannon entropy and, consequently, for the information theory methods (see \textbf{Supplementary Materials}). Figure~\ref{fig:lp} shows that TE (KSG) reliably estimates the `true' transfer entropy but TE (H) significantly underestimates the theoretical values. This is a fundamental flaw that undermines any other advantageous properties of TE (H). Though computed CTIR values match the theoretical values, it is clearly negative in $X\rightarrow Y$ and as such does not reflect the causal structure of the system. Increasing the size of the data $T$ alters the value of TE (H) here, but TE (KSG) remains accurate as $T$ increases. However, in all other simulations, TE (H) is more robust to increasing data size.

The \textbf{Ulam lattice} (UL) chains together unidirectional coupled chaotic Ulam maps. For large $N_L$, the causal influence from $Y$ to $X$ is negligible. UL exhibits synchronisation for $\lambda \approx 0.18,~0.82$, where cause and effect variables are indistinguishable from each other, e.g. the system converges to a two state attractor. As a result, most indices either have values approximately equal to zero or suffer from high variance numerical instabilities. Outside of these regions of synchronisation, the information theoretic methods and regression based indices show reasonable consistency (Figures~\ref{fig:corr} and \ref{fig:hu}). The exception is CTIR, which slowly decreases as $\lambda$ increases for $T = 10^5$, albeit still correctly identifying the direction of information flow. ETE~(H) successfully corrects for the small sample effects that give rise to these spurious positive TE~(H) results in the $Y\rightarrow X$ direction when $T = 10^3$. Both similarity indices fail to identify any causal structure in the UL. For CCM, whilst the net directed index $D_{X\rightarrow Y}$ increases with $\lambda$, it is negative for $\lambda < 0.5$, and so misidentifies the direction of causality. The positive correlations between methods in $i_{Y\rightarrow X}$ for $T = 10^5$ occur due to a very slight peak in value at $\lambda \approx 0.5$ for nearly all methods (except CTIR and EGC).

Synchronisation occurs in the range $\lambda \in [0.7, 1]$ for \textbf{H{\' e}non unidirectional} maps (Figure~\ref{fig:hu}). All indices are consistent and perform reasonably well and we do not observe the noisy fluctuations seen in Ref~\onlinecite{Lungarella2007-vs}, apart from for EGC when $\lambda > 0.7$. It is notable that about half the indices (TE (H), ETE (H), EGC, NLGC, PI) are fairly consistent even as the order of magnitude of the data size $T$ is increased, whilst the values of the other indices (TE (KSG), CTIR, SI, CCM) increase, in some cases tripling in value. There is a strong degree of similarity between all methods for the \textbf{H{\' e}non bidirectional} maps (Figure~\ref{fig:hb0}). In HB (I) simulations, the exceptions to this are NLGC and CCM, though we found that setting a larger number of RBFs resulted in better performance for NLGC. For CCM, the direction of causality is sometimes incorrect and the reason for this is unclear, though this may also be a result of poor parameter choices. We observe expected symmetry in the values of $\lambda$, and synchronisation in the region approximately equal to $\{(\lambda_{xy},\lambda_{yx}):~\lambda_{xy} + \lambda_{yx} > 0.28\}$. There are a small number of points in which numerical instabilities are present in all indices, but the consistency across all indices suggests that these are isolated points in which the system converges to some limit cycle or attractor. There are more differences between methods in HB (NI) results. Significant numerical instabilities occur in EGC, particularly when the system is in a state of synchrony: the region approximately equal to $\{(\lambda_{xy},\lambda_{yx}):~0.05 < \lambda_{xy} < 0.15,~ 0.1 < \lambda_{yx} < 0.28\}$ Outside of this region, EGC is broadly similar to the information theoretic indices, which are highly correlated, and to a slightly lesser degree with SI\textsuperscript{(1)} and the PI. In contrast, NLGC, SI\textsuperscript{(2)} and CCM have unusual results. The first of these is negative almost everywhere (even in a repeat analysis with more RBF kernels) and the latter two are mostly non-negative, and moreover the regions with the most extreme values occur in quite different places in all three.

\subsection*{Computational burden}

An important consideration in selecting a suitable method is any trade-off between performance and computational efficiency. The most significant factor in this is often how the algorithmic cost of each method scales with increasing data size $T$. Table~S.II shows the mean and standard deviations of the time taken to compute each index and simulation. TE (H)/ETE (H) is the fastest in almost all cases, even though this calculation also includes 10 reshuffles and recomputations for ETE (H). CCM, EGC and TE (KSG) are similarly among methods with smaller computational cost. Several methods have extreme values in UL simulations with $T = 10^5$, particularly CTIR and PI, but this is distorted by difficulties in computation when the system is in a synchronised state. We observed a marked difference in computational cost for NLGC when using $k$-means for clustering instead of fuzzy $c$-means and we suggest that $k$-means is more suitable here. Our computation was done in a high performance CPU computing cluster using SkyLake 6140 with 18 core 2.3GHz processors and 384GB of RAM. Although the computational times we report may be slightly faster than on a laptop computer with less processing power, we did not observe any substantial difference when internally comparing run times.

\subsection*{Real-world relevant data issues}

Next, we investigated the sensitivity of all causality indices to a number of modifications mirroring issues that often arise in real-world data. We choose UL with $T = 10^3$ to illustrate the effects of these transformations, as LP is too simplistic a model to give sufficient insight. We keep all simulation parameters and causality index parameters the same. In Table~\ref{tab:transformations}, we summarise the means $\hat{\mu}$ and standard deviations $\hat{\sigma}$ of the directed indices $D_{X\rightarrow Y}$ for each $\lambda$, which are normalised by their deviation from the `base' UL results and averaged over all $\lambda$. This normalisation allows us to compare across methods that take values in different ranges.

\begin{table*}
\caption{Summary of all results from experiments into the effects of data size, scaling, rounding, missing data and Gaussian noise. Taking the Ulam lattice ($T = 10^3$) as a baseline, we recompute the causality indices across 10 independent experimental runs for each $\lambda \in [0,1]$ as before. For each $\lambda$, we compute the mean, $\hat{\mu}$, and standard deviation, $\hat{\sigma}$, of directed indices $D_{X\rightarrow Y}$. We subsequently compute the deviation from the baseline $\mu$ and $\sigma$, reporting the average of these over the $\lambda$ values (excluding the few $\lambda$ values where the system exhibits general synchronisation). We normalise deviations between $\mu$ and $\hat{\mu}$ by the absolute value of $\mu$, with $f(\mu, \hat{\mu}) = \langle\mu - \hat{\mu}\rangle / \langle|\mu|\rangle$ and $g(\sigma, \hat{\sigma}) = \langle\hat{\sigma}\rangle / \langle\sigma\rangle$, where $\langle\cdot\rangle$ is the mean over $\lambda$. If the modified simulation returns the same values as the baseline, then $f = 0$ and $g = 1$. All entries except in the column for the baseline $T = 10^3$ report these $f$ and $g$ values.}
\begin{tabular}{l|r|r|r|rrr|rrr|rr|rrr}
    \noalign{\hrule height 2pt}
    & Baseline & & \multicolumn{1}{c|}{Data size} & \multicolumn{3}{c|}{Scaling} & \multicolumn{3}{c|}{Rounding} & \multicolumn{2}{c|}{Missing data} & \multicolumn{3}{c}{Gaussian noise} \\
    Method & $T = 10^3$ & & \multicolumn{1}{r|}{$T = 10^5$} & Stand. & $D_{10X\rightarrow Y}$ & \multicolumn{1}{r|}{$D_{X\rightarrow 10Y}$} & 1 d.p. & 1 d.p. & \multicolumn{1}{r|}{2 d.p.} & 10\% NA & \multicolumn{1}{r|}{20\% NA} & $\sigma^2_G$ = 0.1 & $\sigma^2_G$ = 1 & $\sigma^2_G$ = 1 \\
    & & & \multicolumn{1}{r|}{} & X,Y & & \multicolumn{1}{r|}{} & X & Y & \multicolumn{1}{r|}{X,Y} & X,Y & \multicolumn{1}{r|}{X,Y} & X,Y & X & Y \\
    \noalign{\hrule height 2pt}
    EGC & $\langle\mu\rangle$ = 0.840 & $f(\mu,\hat{\mu})$ & -0.064 & 0.036 & 0.207 & -0.071 & 0.031 & 0.112 & 0.004 & -0.027 & -0.040 & 0.533 & 0.981 & 0.950 \\
     & $\langle\sigma\rangle$ = 0.021 & $g(\sigma,\hat{\sigma})$ & 0.660 & 1.004 & 1.425 & 0.691 & 0.959 & 0.946 & 0.970 & \textbf{1.023} & \textbf{1.033} & 1.025 & 0.473 & 0.598 \\
    \noalign{\hrule height 1pt}
    NLGC & $\langle\mu\rangle$ = 0.610 & $f(\mu,\hat{\mu})$ & 0.013 & 0.299 & 2.905 & -101.849 & \textbf{0.000} & \textbf{-0.003} & 0.001 & -0.008 & -0.020 & 0.031 & 0.740 & \textbf{-0.007} \\
     & $\langle\sigma\rangle$ = 0.023 & $g(\sigma,\hat{\sigma})$ & 0.089 & 0.608 & 64.469 & 126.734 & \textbf{1.000} & 0.954 & 0.994 & 1.353 & 1.906 & 1.023 & 1.345 & 2.325 \\
    \noalign{\hrule height 1pt}
    PI & $\langle\mu\rangle$ = 0.380 & $f(\mu,\hat{\mu})$ & \textbf{0.002} & 0.293 & 1.576 & -98.727 & 0.950 & -0.951 & -0.016 & \textbf{0.001} & \textbf{0.005} & 0.011 & 0.617 & 0.019 \\
     & $\langle\sigma\rangle$ = 0.032 & $g(\sigma,\hat{\sigma})$ & 0.094 & 0.681 & 69.340 & 66.364 & 0.918 & 1.051 & \textbf{1.001} & 1.214 & 1.511 & \textbf{1.007} & 2.041 & 2.120 \\
    \noalign{\hrule height 1pt}
    TE (H) & $\langle\mu\rangle$ = 0.675 & $f(\mu,\hat{\mu})$ & -0.158 & \textbf{0.000} & \textbf{0.000} & \textbf{0.000} & 0.011 & 0.030 & \textbf{0.000} & 0.071 & 0.159 & 0.026 & 0.786 & 0.731 \\
     & $\langle\sigma\rangle$ = 0.019 & $g(\sigma,\hat{\sigma})$ & 0.085 & \textbf{1.000} & \textbf{1.000} & \textbf{1.000} & 1.004 & \textbf{0.994} & 1.013 & 1.313 & 1.633 & 1.112 & \textbf{1.015} & 0.920 \\
    \noalign{\hrule height 1pt}
    ETE (H) & $\langle\mu\rangle$ = 0.674 & $f(\mu,\hat{\mu})$ & -0.158 & \textbf{0.000} & \textbf{0.000} & \textbf{0.000} & 0.014 & 0.026 & 0.000 & 0.075 & 0.155 & 0.026 & 0.748 & 0.774 \\
     & $\langle\sigma\rangle$ = 0.019 & $g(\sigma,\hat{\sigma})$ & 0.083 & \textbf{1.000} & \textbf{1.000} & \textbf{1.000} & 0.992 & 0.994 & 1.009 & 1.296 & 1.634 & 1.109 & 0.947 & 0.854 \\
    \noalign{\hrule height 1pt}
    TE (KSG) & $\langle\mu\rangle$ = 1.509 & $f(\mu,\hat{\mu})$ & -1.348 & \textbf{0.000} & 0.269 & 0.609 & 0.095 & -0.134 & -0.029 & 0.111 & 0.216 & 0.306 & 0.841 & 0.863 \\
     & $\langle\sigma\rangle$ = 0.025 & $g(\sigma,\hat{\sigma})$ & 0.120 & 1.071 & 0.869 & 1.026 & 1.232 & 1.320 & 1.042 & 1.197 & 1.430 & 1.028 & 1.017 & \textbf{0.962} \\
    \noalign{\hrule height 1pt}
    CTIR & $\langle\mu\rangle$ = 0.462 & $f(\mu,\hat{\mu})$ & -1.226 & \textbf{0.000} & 0.128 & 0.713 & 0.299 & -0.355 & -0.026 & 0.083 & 0.161 & 0.273 & 0.826 & 0.848 \\
     & $\langle\sigma\rangle$ = 0.014 & $g(\sigma,\hat{\sigma})$ & 0.110 & 1.016 & 0.898 & 0.920 & 1.159 & 1.209 & 1.042 & 1.076 & 1.258 & 0.958 & 0.942 & 0.872 \\
    \noalign{\hrule height 1pt}
    SI$^{(1)}$ & $\langle\mu\rangle$ = 0.001 & $f(\mu,\hat{\mu})$ & 0.015 & \textbf{0.000} & \textbf{0.000} & \textbf{0.000} & 0.549 & -0.556 & 0.005 & -0.013 & 0.018 & -0.007 & \textbf{-0.061} & 0.066 \\
     & $\langle\sigma\rangle$ = 0.029 & $g(\sigma,\hat{\sigma})$ & 0.097 & \textbf{1.000} & \textbf{1.000} & \textbf{1.000} & 1.363 & 1.355 & 1.053 & 1.084 & 1.219 & 0.895 & 0.705 & 0.693 \\
    \noalign{\hrule height 1pt}
    SI$^{(2)}$ & $\langle\mu\rangle$ = 0.000 & $f(\mu,\hat{\mu})$ & 0.029 & \textbf{0.000} & \textbf{0.000} & \textbf{0.000} & -3.105 & 3.121 & -0.001 & -0.037 & 0.017 & \textbf{0.000} & -8.256 & 7.736 \\
     & $\langle\sigma\rangle$ = 0.000 & $g(\sigma,\hat{\sigma})$ & \textbf{0.000} & \textbf{1.000} & \textbf{1.000} & \textbf{1.000} & 1.394 & 1.399 & 1.074 & 1.666 & 2.630 & 0.968 & 2.334 & 2.018 \\
    \noalign{\hrule height 1pt}
    CCM & $\langle\mu\rangle$ = 0.001 & $f(\mu,\hat{\mu})$ & 0.031 & \textbf{0.000} & \textbf{0.000} & \textbf{0.000} & -1.249 & 1.289 & -0.005 & -0.009 & -0.025 & 0.013 & 0.151 & -0.075 \\
     & $\langle\sigma\rangle$ = 0.047 & $g(\sigma,\hat{\sigma})$ & 0.103 & \textbf{1.000} & \textbf{1.000} & \textbf{1.000} & 1.115 & 1.105 & 0.740 & 1.090 & 1.250 & 1.010 & 0.944 & 0.959 \\
    \noalign{\hrule height 2pt}
\end{tabular}
\label{tab:transformations}
\end{table*}

\subsubsection{Data availability}

Many of the indices, with the exception of TE (KSG) and CTIR, remain consistent with increasing data size $T$, whilst at the same time exhibit decreasing variance. These results reinforce the similar observations in HU maps. The large increases in the value of the two methods mentioned is concerning and represent a drawback of both methods that should be acknowledged in applications of these approaches. It is unclear whether there is convergence to some `correct' value as the amount of data increases or whether both are unbounded as $T \rightarrow \infty$, but initial computations do not support the former (not shown). Though the $D_{X\rightarrow Y}$ values from both transfer entropy methods are highly correlated, they are both estimates of the same quantity and it is difficult to reconcile their different magnitudes, particularly as we have already seen significant underestimation in TE (H) for LP simulations.

\subsubsection{\textit{Standardisation and scaling}}

In the second set of experiments, we perform three tests: standardising both series by their sample mean and standard deviation in the first, and separately scaling each unstandardised time series by a factor of $10$ (Figure~S2). For the Ulam lattice system, sample means for both $X$ and $Y$ are typically between 0.4 and 0.7 and standard deviations are both approximately equal to 1.2 (except when the system is in synchrony). Several methods are invariant under linear scaling or shifting of the original series $X$ and $Y$, including cross mapping approaches. Information theoretic measures are also invariant in theory, but the KSG algorithm, based on $k$-nearest neighbours, does not retain this property. Similarly, EGC relies on a neighbourhood size parameter, and scaling the data without changing this parameter accordingly can result in insufficient points available for the locally linear regressions, as is observed when either $X$ or $Y$ is scaled by 10. The directed index for both NLGC and PI has vastly inflated magnitude when $Y$ is scaled by 10. With this in mind, we recommend standardisation or normalisation of the data before employing these methods.

\subsubsection{\textit{Rounding error and missing data}}

We perform three experiments to investigate rounding error, first rounding each time series separately to 1 decimal place and then rounding both to 2 decimal places (Figure~S3). TE and both GC extensions have similar performance to the baseline in all cases, whilst CCM suffers the most. In two experiments with missing data of 10\% and 20\%, all methods appear robust to this.

\subsubsection{\textit{Noisy data}}

In the case of the earlier LP simulations, Gaussian noise forms an integral component of the system itself and the theoretical expression for TE shows that this depends only on the ratio of the variances $\sigma_x / \sigma_y$ (see \textbf{Supplementary Material}). However, this noise is inherent in the simulation process (i.e. it does not arise in observation of the system). In our UL experiments, we added additional Gaussian noise \textit{after} simulation. The inclusion of this `observation' noise does not alter the state of the system or the information flow between variables but it does obscure the causal structure. In the first of these experiments (Figure~S3), in which we added small variance Gaussian noise ($\sigma_{G} = 0.1$), the amplitude of this noise is an order of magnitude less than the original UL values and the inclusion of this noise has a small effect for all indices. In the latter experiments, we added Gaussian noise (with $\sigma_G = 1$) to each variable individually and the effect is more pronounced. NLGC performs best in general and appears very resilient to noise added to $Y$ (effect variable), though it drops slightly in value when Gaussian noise is added to $X$ (cause variable). It is interesting that the two SI have quite different results, with SI$^{(1)}$ more robust to noise, though both methods are not in general able to successfully identify the direction of causality.

\section*{Discussion}

In-depth comparative studies of this kind are relatively rare in the mathematical literature (examples include Refs.~\onlinecite{Papana2013-ul,Cutts2014-ow}), particularly in evaluating performance of methods for estimating a concept, such as causality, that does not have a consistent, fundamental mathematical definition. Even without this, causal inference has a huge importance in how we can model, predict and exploit real-world applications from many scientific disciplines. Asymmetric bivariate causal inference is the first key step to providing this insight into interactions between components in complex networks. In the context of causality indices, review papers \cite{Hlavackova-Schindler2007-lm, Eichler2013-pb, Papana2013-ul} have previously had a narrower focus in some manner, for example on only one group of methods or on a few bivariate methods and their multivariate extensions. We follow the template of Ref~\onlinecite{Lungarella2007-vs} in reviewing methods drawn from diverse mathematical foundations, but we extend this review with additional methods and crucially we investigate the impact of common issues that are relevant to real-world data. In reproducing and updating their work, we are also able to resolve some computational stability issues and comment on the computational costs of each method, whilst we also make our code publicly available for other researchers to develop further. 

\subsection*{Further work}

A primary concern in causality inference is the difficulties with model misspecification, specifically causal identification in multivariate systems. Omission of confounding variables can create spurious false-positive causal relationships. There may also be redundancy across multiple variables that provide similar information to the effect variable or sets of variables that interact synergistically such that their combined causal influence is greater than the `sum of their parts'. These are key concerns outlined in Ref~\onlinecite{Eichler2013-pb} and, consequently, results from bivariate indices cannot be definitively interpreted as the existence of a fundamental direct causal relationship between two variables \cite{Arnhold1999-si}. A key avenue for further work is to advance this analysis beyond a bivariate setting by including possible confounding variables, in line with conditional extensions to Granger causality \cite{Geweke1984-er, Chen2004-vu, Siggiridou2016-yj, Guo2008-nx} and transfer entropy \cite{Lizier2014-xt}. Recent work with graphical models of multivariate systems \cite{Runge2018-dk} is an important step towards high-dimensional causal identification.

Separate univariate embedding is not without some limitations and is not necessarily the optimal multivariate embedding. Aside from in Ref~\onlinecite{Vlachos2010-ta}, mixed embeddings are as yet uncommon in causality estimation. There is not yet a theoretical framework for longitudinal data that is recorded non-simultaneous and irregularly. Often, a typical workflow for such data involves pre-processing to transform the data into a multivariate series with constant time intervals. However, many imputation methods result in significant and poorly quantified biases in information content and flow, which inevitably propagate through to estimates of causality, and more work is needed to explicitly factor this into a causal inference framework.

\subsection*{Summary and recommendations}

Each causality index has strengths and weaknesses, and there is no single method whose all-round performance exceeds all others. Transfer entropy and Granger causality have long been regarded as the leading methods for systems that contain a small number of variables, and these have had wide applications \cite{Granger1969-pv, Geweke1984-er,Seth2015-rp}. Transfer entropy has the distinct advantage that it is built upon the principles of Shannon entropy in a well-established and universal information theoretic framework. It performs solidly throughout, though there is some tension between algorithms for TE, with the estimates rarely in complete agreement. We have shown that a histogram fixed partition approach is biased even in the simplest model, despite TE (H) having general consistency and computationally efficiency. Therefore, we recommend the KSG algorithm for transfer entropy computation, unless perhaps data is extremely scarce ($T < 10^3$). However, there are some unanswered concerns about TE (KSG), particularly that it appears to increase in magnitude as more data is available. TE (KSG) also suffers in performance when data is unequally scaled, due to the resultant difficulties with identifying unique nearest neighbours. CTIR, whilst sometimes not wholly dissimilar in value from TE, did not seem to offer any obvious advantage to compensate for its much higher computational cost or occasional unusual behaviour. Vanilla GC is widely favoured but has restrictive assumptions and is ill-suited to complex nonlinear problems. Of the two nonlinear extensions to Granger causality, Lungarella \textit{et al.} \cite{Lungarella2007-vs} appear to prefer EGC. Some of the computational challenges and numerical instability that they experienced with NLGC may have been a result of their choice of a fuzzy $c$-means for determining RBF kernels, and alternate parameter choices appear to resolve some of their concerns. We find that NLGC is one of the most robust methods to rounding error, missing data and Gaussian noise. They rightly note that "If the rank of the data is small, kernel based methods tend to overfit" \cite{Lungarella2007-vs}, but we did not observe any issues with this in our simulation experiments. Predictability improvement (PI) likewise performed solidly, and has a slight advantage amongst the regression based indices in that it that it is perhaps less reliant on parameter choices. Finally, dynamical systems theory offers a different insight into causal inference that should not be readily dismissed despite our mixed results here, even though the deterministic simulation models appeared to be well-suited to the underlying theory. Convergent cross mapping is a more recent and popular method, and this offered a broad improvement on the similarity indices (SI), which did not consistently identify the strength or direction of causality. However, CCM too did not always manage to determine the correct causal flow in our simulations. 

We have highlighted the value of a standardisation pre-processing step in in avoiding algorithmic issues, which is also important in comparing results from different data for each method. Rounding error gives rise to practical issues within the implementation of several of the algorithms. For instance, in $k$-nearest neighbour approaches it is typically assumed that the distances between pairs of points are unique and not discrete. Subsequent edge cases can be treated by adding random noise with low amplitude to the data before estimating the causal relationships \cite{Kraskov2004-sf}, though propagation of this noise to final estimates is something that should be analysed. Likewise, many existing implementations of the methods are not equipped to handle missing data (e.g. Refs.~\onlinecite{Lizier2014-xt,Park2020-su}). We believe this is broadly straightforward to implement across all indices, as it can be handled exclusively within the time-delay embedding vector step, by performing an embedding and then removing any embedding vectors missing at least one component. As M{\" o}nster \textit{et al.} \cite{Monster2016-wy} put it, "Noise in real-world data is ubiquitous, the inclusion of noise in model investigations has been largely ignored". Added Gaussian noise leads to the biggest changes in value for most methods, particularly noise at observation in the causal variable. However, provided the magnitude of noise is small compared to the values themselves, all methods perform adequately.

On the basis of this work, we conclude that the strongest choice for identifying and quantifying bivariate causal relationships is, in our view, either \textbf{transfer entropy (KSG)} or \textbf{nonlinear Granger causality}. Predictability improvement is a reasonable alternative and perhaps the next best candidate. A more cautious approach may involve using more than one method, from different theoretical backgrounds. Where possible, it is advantageous to identify a base case for the system, which subsequent results can be reliably compared against. For new methodologies, we recommend investigation into the real-world issues we have discussed.

\section*{Code and data availability}

Our code is openly available at the GitHub repository \url{https://github.com/tedinburgh/causality-review} and Ref~\onlinecite{Edinburgh2021-ha}. The data that support the findings of this study are openly available at the same repository. A CODECHECK certificate is available confirming that the computations underlying this article could be independently executed: doi.org/10.5281/zenodo.4720843.

Existing open-access code for some indices include repositories for information theory and transfer entropy: IDTxl \cite{Lizier2014-xt} v1.1, PyIF \cite{Brunner2019-ip}; and for convergent cross mapping: pyEDM \cite{Park2020-su} v1.7.4. We also adapted fuzzy $c$-means code based on Ref~\onlinecite{NourJamalElDin2020-su}. We checked our results for transfer entropy and convergent cross mapping against those from IDTxl and pyEDM respectively. All code in our repository and in these others is Python.

\section*{Supplementary Materials}

Our supplementary materials contains additional tables and figures. Tables~S.I and S.II show full parameter choices and computational time requirements of each method respectively. Figures~S1-3 show the results of real-world relevant transformation experiments. The supplementary materials also contain theoretical results for information theoretic measures in the linear process simulation.

\begin{acknowledgments}
TE is funded by Engineering and Physical Sciences Research Council (EPSRC) National Productivity Investment Fund (NPIF) EP/S515334/1, reference 2089662. A CC BY or equivalent licence is applied to the AAM arising from this submission: this manuscript is the AAM.

We would like to acknowledge Marcel Stimberg and Daniel N{\"u}st for their work with the CODECHECK. We found several existing open-source code repositories, listed above in \textbf{Code and data availability}. It was insightful to view and test these packages, though we still decided to develop our own code for these methods. In addition, we appreciate advice from George Sugihara on use of CCM (pyEDM) in email correspondence. We also acknowledge the Python community for core packages that this work depends upon, including ipython \cite{Perez2007-lt} v7.16.1, matplotlib \cite{Hunter2007-vy} v3.3.2, numpy \cite{Harris2020-od} v1.18.5, palettable \cite{Davis2012-ip} v3.3.0, pandas \cite{McKinney2010-ey} v1.0.5, python \cite{VanRossum1995-aa} v3.8.3, scikit-learn \cite{Pedregosa2011-bt} v0.23.1, scipy \cite{Virtanen2020-db} v1.5.0 and statsmodels \cite{Seabold2010-ab} v0.11.1.
\end{acknowledgments}

\bibliography{references}

\providecommand{\noopsort}[1]{}\providecommand{\singleletter}[1]{#1}%
\begin{thebibliography}{56}%
\makeatletter
\providecommand \@ifxundefined [1]{%
 \@ifx{#1\undefined}
}%
\providecommand \@ifnum [1]{%
 \ifnum #1\expandafter \@firstoftwo
 \else \expandafter \@secondoftwo
 \fi
}%
\providecommand \@ifx [1]{%
 \ifx #1\expandafter \@firstoftwo
 \else \expandafter \@secondoftwo
 \fi
}%
\providecommand \natexlab [1]{#1}%
\providecommand \enquote  [1]{``#1''}%
\providecommand \bibnamefont  [1]{#1}%
\providecommand \bibfnamefont [1]{#1}%
\providecommand \citenamefont [1]{#1}%
\providecommand \href@noop [0]{\@secondoftwo}%
\providecommand \href [0]{\begingroup \@sanitize@url \@href}%
\providecommand \@href[1]{\@@startlink{#1}\@@href}%
\providecommand \@@href[1]{\endgroup#1\@@endlink}%
\providecommand \@sanitize@url [0]{\catcode `\\12\catcode `\$12\catcode
  `\&12\catcode `\#12\catcode `\^12\catcode `\_12\catcode `\%12\relax}%
\providecommand \@@startlink[1]{}%
\providecommand \@@endlink[0]{}%
\providecommand \url  [0]{\begingroup\@sanitize@url \@url }%
\providecommand \@url [1]{\endgroup\@href {#1}{\urlprefix }}%
\providecommand \urlprefix  [0]{URL }%
\providecommand \Eprint [0]{\href }%
\providecommand \doibase [0]{http://dx.doi.org/}%
\providecommand \selectlanguage [0]{\@gobble}%
\providecommand \bibinfo  [0]{\@secondoftwo}%
\providecommand \bibfield  [0]{\@secondoftwo}%
\providecommand \translation [1]{[#1]}%
\providecommand \BibitemOpen [0]{}%
\providecommand \bibitemStop [0]{}%
\providecommand \bibitemNoStop [0]{.\EOS\space}%
\providecommand \EOS [0]{\spacefactor3000\relax}%
\providecommand \BibitemShut  [1]{\csname bibitem#1\endcsname}%
\let\auto@bib@innerbib\@empty
\bibitem [{\citenamefont {Granger}(1969{\natexlab{a}})}]{Granger1969-bj}%
  \BibitemOpen
  \bibfield  {author} {\bibinfo {author} {\bibfnamefont {C.~W.~J.}\
  \bibnamefont {Granger}},\ }\bibfield  {title} {\enquote {\bibinfo {title}
  {Investigating causal relations by econometric models and cross-spectral
  methods},}\ }\href@noop {} {\bibfield  {journal} {\bibinfo  {journal}
  {Econometrica}\ }\textbf {\bibinfo {volume} {37}},\ \bibinfo {pages}
  {424--438} (\bibinfo {year} {1969}{\natexlab{a}})}\BibitemShut {NoStop}%
\bibitem [{\citenamefont {Runge}(2018)}]{Runge2018-dk}%
  \BibitemOpen
  \bibfield  {author} {\bibinfo {author} {\bibfnamefont {J.}~\bibnamefont
  {Runge}},\ }\bibfield  {title} {\enquote {\bibinfo {title} {Causal network
  reconstruction from time series: From theoretical assumptions to practical
  estimation},}\ }\href@noop {} {\bibfield  {journal} {\bibinfo  {journal}
  {Chaos}\ }\textbf {\bibinfo {volume} {28}},\ \bibinfo {pages} {075310}
  (\bibinfo {year} {2018})}\BibitemShut {NoStop}%
\bibitem [{\citenamefont {Eichler}(2012)}]{Eichler2012-tw}%
  \BibitemOpen
  \bibfield  {author} {\bibinfo {author} {\bibfnamefont {M.}~\bibnamefont
  {Eichler}},\ }\bibfield  {title} {\enquote {\bibinfo {title} {Graphical
  modelling of multivariate time series},}\ }\href@noop {} {\bibfield
  {journal} {\bibinfo  {journal} {Probab. Theory Related Fields}\ }\textbf
  {\bibinfo {volume} {153}},\ \bibinfo {pages} {233--268} (\bibinfo {year}
  {2012})}\BibitemShut {NoStop}%
\bibitem [{\citenamefont {Aldrich}(1995)}]{Aldrich1995-iz}%
  \BibitemOpen
  \bibfield  {author} {\bibinfo {author} {\bibfnamefont {J.}~\bibnamefont
  {Aldrich}},\ }\bibfield  {title} {\enquote {\bibinfo {title} {Correlations
  genuine and spurious in {Pearson} and {Yule}},}\ }\href@noop {} {\bibfield
  {journal} {\bibinfo  {journal} {Stat. Sci.}\ }\textbf {\bibinfo {volume}
  {10}},\ \bibinfo {pages} {364--376} (\bibinfo {year} {1995})}\BibitemShut
  {NoStop}%
\bibitem [{\citenamefont {Sugihara}\ \emph {et~al.}(2012)\citenamefont
  {Sugihara}, \citenamefont {May}, \citenamefont {Ye}, \citenamefont {Hsieh},
  \citenamefont {Deyle}, \citenamefont {Fogarty},\ and\ \citenamefont
  {Munch}}]{Sugihara2012-lv}%
  \BibitemOpen
  \bibfield  {author} {\bibinfo {author} {\bibfnamefont {G.}~\bibnamefont
  {Sugihara}}, \bibinfo {author} {\bibfnamefont {R.}~\bibnamefont {May}},
  \bibinfo {author} {\bibfnamefont {H.}~\bibnamefont {Ye}}, \bibinfo {author}
  {\bibfnamefont {C.-H.}\ \bibnamefont {Hsieh}}, \bibinfo {author}
  {\bibfnamefont {E.}~\bibnamefont {Deyle}}, \bibinfo {author} {\bibfnamefont
  {M.}~\bibnamefont {Fogarty}}, \ and\ \bibinfo {author} {\bibfnamefont
  {S.}~\bibnamefont {Munch}},\ }\bibfield  {title} {\enquote {\bibinfo {title}
  {Detecting causality in complex ecosystems},}\ }\href@noop {} {\bibfield
  {journal} {\bibinfo  {journal} {Science}\ }\textbf {\bibinfo {volume}
  {338}},\ \bibinfo {pages} {496--500} (\bibinfo {year} {2012})}\BibitemShut
  {NoStop}%
\bibitem [{\citenamefont {Sims}(1972)}]{Sims1972-gj}%
  \BibitemOpen
  \bibfield  {author} {\bibinfo {author} {\bibfnamefont {C.~A.}\ \bibnamefont
  {Sims}},\ }\bibfield  {title} {\enquote {\bibinfo {title} {Money, income, and
  causality},}\ }\href@noop {} {\bibfield  {journal} {\bibinfo  {journal} {Am.
  Econ. Rev.}\ }\textbf {\bibinfo {volume} {62}},\ \bibinfo {pages} {540--552}
  (\bibinfo {year} {1972})}\BibitemShut {NoStop}%
\bibitem [{\citenamefont {Schreiber}(2000)}]{Schreiber2000-hz}%
  \BibitemOpen
  \bibfield  {author} {\bibinfo {author} {\bibfnamefont {T.}~\bibnamefont
  {Schreiber}},\ }\bibfield  {title} {\enquote {\bibinfo {title} {Measuring
  information transfer},}\ }\href@noop {} {\bibfield  {journal} {\bibinfo
  {journal} {Phys. Rev. Lett.}\ }\textbf {\bibinfo {volume} {85}},\ \bibinfo
  {pages} {461--464} (\bibinfo {year} {2000})}\BibitemShut {NoStop}%
\bibitem [{\citenamefont {Granger}(1969{\natexlab{b}})}]{Granger1969-pv}%
  \BibitemOpen
  \bibfield  {author} {\bibinfo {author} {\bibfnamefont {C.~W.~J.}\
  \bibnamefont {Granger}},\ }\bibfield  {title} {\enquote {\bibinfo {title}
  {Investigating causal relations by econometric models and cross-spectral
  methods},}\ }\href@noop {} {\bibfield  {journal} {\bibinfo  {journal}
  {Econometrica}\ }\textbf {\bibinfo {volume} {37}},\ \bibinfo {pages}
  {424--438} (\bibinfo {year} {1969}{\natexlab{b}})}\BibitemShut {NoStop}%
\bibitem [{\citenamefont {Geweke}(1984)}]{Geweke1984-er}%
  \BibitemOpen
  \bibfield  {author} {\bibinfo {author} {\bibfnamefont {J.}~\bibnamefont
  {Geweke}},\ }\bibfield  {title} {\enquote {\bibinfo {title} {Inference and
  causality in economic time series models},}\ }in\ \href@noop {} {\emph
  {\bibinfo {booktitle} {Handbook of Econometrics}}},\ Vol.~\bibinfo {volume}
  {2}\ (\bibinfo  {publisher} {Elsevier},\ \bibinfo {year} {1984})\ pp.\
  \bibinfo {pages} {1101--1144}\BibitemShut {NoStop}%
\bibitem [{\citenamefont {Zhang}\ \emph {et~al.}(2011)\citenamefont {Zhang},
  \citenamefont {Lee}, \citenamefont {Wang}, \citenamefont {Li}, \citenamefont
  {Pei}, \citenamefont {Zhang},\ and\ \citenamefont {An}}]{Zhang2011-fc}%
  \BibitemOpen
  \bibfield  {author} {\bibinfo {author} {\bibfnamefont {D.~D.}\ \bibnamefont
  {Zhang}}, \bibinfo {author} {\bibfnamefont {H.~F.}\ \bibnamefont {Lee}},
  \bibinfo {author} {\bibfnamefont {C.}~\bibnamefont {Wang}}, \bibinfo {author}
  {\bibfnamefont {B.}~\bibnamefont {Li}}, \bibinfo {author} {\bibfnamefont
  {Q.}~\bibnamefont {Pei}}, \bibinfo {author} {\bibfnamefont {J.}~\bibnamefont
  {Zhang}}, \ and\ \bibinfo {author} {\bibfnamefont {Y.}~\bibnamefont {An}},\
  }\bibfield  {title} {\enquote {\bibinfo {title} {The causality analysis of
  climate change and large-scale human crisis},}\ }\href@noop {} {\bibfield
  {journal} {\bibinfo  {journal} {Proc. Natl. Acad. Sci. U. S. A.}\ }\textbf
  {\bibinfo {volume} {108}},\ \bibinfo {pages} {17296--17301} (\bibinfo {year}
  {2011})}\BibitemShut {NoStop}%
\bibitem [{\citenamefont {Runge}\ \emph
  {et~al.}(2019{\natexlab{a}})\citenamefont {Runge}, \citenamefont {Nowack},
  \citenamefont {Kretschmer}, \citenamefont {Flaxman},\ and\ \citenamefont
  {Sejdinovic}}]{Runge2019-zq}%
  \BibitemOpen
  \bibfield  {author} {\bibinfo {author} {\bibfnamefont {J.}~\bibnamefont
  {Runge}}, \bibinfo {author} {\bibfnamefont {P.}~\bibnamefont {Nowack}},
  \bibinfo {author} {\bibfnamefont {M.}~\bibnamefont {Kretschmer}}, \bibinfo
  {author} {\bibfnamefont {S.}~\bibnamefont {Flaxman}}, \ and\ \bibinfo
  {author} {\bibfnamefont {D.}~\bibnamefont {Sejdinovic}},\ }\bibfield  {title}
  {\enquote {\bibinfo {title} {Detecting and quantifying causal associations in
  large nonlinear time series datasets},}\ }\href@noop {} {\bibfield  {journal}
  {\bibinfo  {journal} {Sci Adv}\ }\textbf {\bibinfo {volume} {5}},\ \bibinfo
  {pages} {eaau4996} (\bibinfo {year} {2019}{\natexlab{a}})}\BibitemShut
  {NoStop}%
\bibitem [{\citenamefont {Runge}\ \emph
  {et~al.}(2019{\natexlab{b}})\citenamefont {Runge}, \citenamefont {Bathiany},
  \citenamefont {Bollt}, \citenamefont {Camps-Valls}, \citenamefont {Coumou},
  \citenamefont {Deyle}, \citenamefont {Glymour}, \citenamefont {Kretschmer},
  \citenamefont {Mahecha}, \citenamefont {Mu{\~n}oz-Mar{\'\i}}, \citenamefont
  {van Nes}, \citenamefont {Peters}, \citenamefont {Quax}, \citenamefont
  {Reichstein}, \citenamefont {Scheffer}, \citenamefont {Sch{\"o}lkopf},
  \citenamefont {Spirtes}, \citenamefont {Sugihara}, \citenamefont {Sun},
  \citenamefont {Zhang},\ and\ \citenamefont {Zscheischler}}]{Runge2019-nm}%
  \BibitemOpen
  \bibfield  {author} {\bibinfo {author} {\bibfnamefont {J.}~\bibnamefont
  {Runge}}, \bibinfo {author} {\bibfnamefont {S.}~\bibnamefont {Bathiany}},
  \bibinfo {author} {\bibfnamefont {E.}~\bibnamefont {Bollt}}, \bibinfo
  {author} {\bibfnamefont {G.}~\bibnamefont {Camps-Valls}}, \bibinfo {author}
  {\bibfnamefont {D.}~\bibnamefont {Coumou}}, \bibinfo {author} {\bibfnamefont
  {E.}~\bibnamefont {Deyle}}, \bibinfo {author} {\bibfnamefont
  {C.}~\bibnamefont {Glymour}}, \bibinfo {author} {\bibfnamefont
  {M.}~\bibnamefont {Kretschmer}}, \bibinfo {author} {\bibfnamefont {M.~D.}\
  \bibnamefont {Mahecha}}, \bibinfo {author} {\bibfnamefont {J.}~\bibnamefont
  {Mu{\~n}oz-Mar{\'\i}}}, \bibinfo {author} {\bibfnamefont {E.~H.}\
  \bibnamefont {van Nes}}, \bibinfo {author} {\bibfnamefont {J.}~\bibnamefont
  {Peters}}, \bibinfo {author} {\bibfnamefont {R.}~\bibnamefont {Quax}},
  \bibinfo {author} {\bibfnamefont {M.}~\bibnamefont {Reichstein}}, \bibinfo
  {author} {\bibfnamefont {M.}~\bibnamefont {Scheffer}}, \bibinfo {author}
  {\bibfnamefont {B.}~\bibnamefont {Sch{\"o}lkopf}}, \bibinfo {author}
  {\bibfnamefont {P.}~\bibnamefont {Spirtes}}, \bibinfo {author} {\bibfnamefont
  {G.}~\bibnamefont {Sugihara}}, \bibinfo {author} {\bibfnamefont
  {J.}~\bibnamefont {Sun}}, \bibinfo {author} {\bibfnamefont {K.}~\bibnamefont
  {Zhang}}, \ and\ \bibinfo {author} {\bibfnamefont {J.}~\bibnamefont
  {Zscheischler}},\ }\bibfield  {title} {\enquote {\bibinfo {title} {Inferring
  causation from time series in earth system sciences},}\ }\href@noop {}
  {\bibfield  {journal} {\bibinfo  {journal} {Nat. Commun.}\ }\textbf {\bibinfo
  {volume} {10}},\ \bibinfo {pages} {2553} (\bibinfo {year}
  {2019}{\natexlab{b}})}\BibitemShut {NoStop}%
\bibitem [{\citenamefont {Gray}\ \emph {et~al.}(1989)\citenamefont {Gray},
  \citenamefont {K{\"o}nig}, \citenamefont {Engel},\ and\ \citenamefont
  {Singer}}]{Gray1989-zx}%
  \BibitemOpen
  \bibfield  {author} {\bibinfo {author} {\bibfnamefont {C.~M.}\ \bibnamefont
  {Gray}}, \bibinfo {author} {\bibfnamefont {P.}~\bibnamefont {K{\"o}nig}},
  \bibinfo {author} {\bibfnamefont {A.~K.}\ \bibnamefont {Engel}}, \ and\
  \bibinfo {author} {\bibfnamefont {W.}~\bibnamefont {Singer}},\ }\bibfield
  {title} {\enquote {\bibinfo {title} {Oscillatory responses in cat visual
  cortex exhibit inter-columnar synchronization which reflects global stimulus
  properties},}\ }\href@noop {} {\bibfield  {journal} {\bibinfo  {journal}
  {Nature}\ }\textbf {\bibinfo {volume} {338}},\ \bibinfo {pages} {334--337}
  (\bibinfo {year} {1989})}\BibitemShut {NoStop}%
\bibitem [{\citenamefont {Seth}, \citenamefont {Barrett},\ and\ \citenamefont
  {Barnett}(2015)}]{Seth2015-rp}%
  \BibitemOpen
  \bibfield  {author} {\bibinfo {author} {\bibfnamefont {A.~K.}\ \bibnamefont
  {Seth}}, \bibinfo {author} {\bibfnamefont {A.~B.}\ \bibnamefont {Barrett}}, \
  and\ \bibinfo {author} {\bibfnamefont {L.}~\bibnamefont {Barnett}},\
  }\bibfield  {title} {\enquote {\bibinfo {title} {Granger causality analysis
  in neuroscience and neuroimaging},}\ }\href@noop {} {\bibfield  {journal}
  {\bibinfo  {journal} {J. Neurosci.}\ }\textbf {\bibinfo {volume} {35}},\
  \bibinfo {pages} {3293--3297} (\bibinfo {year} {2015})}\BibitemShut {NoStop}%
\bibitem [{\citenamefont {Hlav{\'a}{\v c}kov{\'a}-Schindler}\ \emph
  {et~al.}(2007)\citenamefont {Hlav{\'a}{\v c}kov{\'a}-Schindler},
  \citenamefont {Palu{\v s}}, \citenamefont {Vejmelka},\ and\ \citenamefont
  {Bhattacharya}}]{Hlavackova-Schindler2007-lm}%
  \BibitemOpen
  \bibfield  {author} {\bibinfo {author} {\bibfnamefont {K.}~\bibnamefont
  {Hlav{\'a}{\v c}kov{\'a}-Schindler}}, \bibinfo {author} {\bibfnamefont
  {M.}~\bibnamefont {Palu{\v s}}}, \bibinfo {author} {\bibfnamefont
  {M.}~\bibnamefont {Vejmelka}}, \ and\ \bibinfo {author} {\bibfnamefont
  {J.}~\bibnamefont {Bhattacharya}},\ }\bibfield  {title} {\enquote {\bibinfo
  {title} {Causality detection based on information-theoretic approaches in
  time series analysis},}\ }\href@noop {} {\bibfield  {journal} {\bibinfo
  {journal} {Phys. Rep.}\ }\textbf {\bibinfo {volume} {441}},\ \bibinfo {pages}
  {1--46} (\bibinfo {year} {2007})}\BibitemShut {NoStop}%
\bibitem [{\citenamefont {Eichler}(2013)}]{Eichler2013-pb}%
  \BibitemOpen
  \bibfield  {author} {\bibinfo {author} {\bibfnamefont {M.}~\bibnamefont
  {Eichler}},\ }\bibfield  {title} {\enquote {\bibinfo {title} {Causal
  inference with multiple time series: principles and problems},}\ }\href@noop
  {} {\bibfield  {journal} {\bibinfo  {journal} {Philos. Trans. A Math. Phys.
  Eng. Sci.}\ }\textbf {\bibinfo {volume} {371}},\ \bibinfo {pages} {20110613}
  (\bibinfo {year} {2013})}\BibitemShut {NoStop}%
\bibitem [{\citenamefont {Papana}\ \emph {et~al.}(2013)\citenamefont {Papana},
  \citenamefont {Kyrtsou}, \citenamefont {Kugiumtzis},\ and\ \citenamefont
  {Diks}}]{Papana2013-ul}%
  \BibitemOpen
  \bibfield  {author} {\bibinfo {author} {\bibfnamefont {A.}~\bibnamefont
  {Papana}}, \bibinfo {author} {\bibfnamefont {C.}~\bibnamefont {Kyrtsou}},
  \bibinfo {author} {\bibfnamefont {D.}~\bibnamefont {Kugiumtzis}}, \ and\
  \bibinfo {author} {\bibfnamefont {C.}~\bibnamefont {Diks}},\ }\bibfield
  {title} {\enquote {\bibinfo {title} {Simulation study of direct causality
  measures in multivariate time series},}\ }\href@noop {} {\bibfield  {journal}
  {\bibinfo  {journal} {Entropy}\ }\textbf {\bibinfo {volume} {15}},\ \bibinfo
  {pages} {2635--2661} (\bibinfo {year} {2013})}\BibitemShut {NoStop}%
\bibitem [{\citenamefont {Palachy}(2019)}]{Palachy2019-kx}%
  \BibitemOpen
  \bibfield  {author} {\bibinfo {author} {\bibfnamefont {S.}~\bibnamefont
  {Palachy}},\ }\href@noop {} {\enquote {\bibinfo {title} {Inferring causality
  in time series data - towards data science},}\ }\bibinfo {howpublished}
  {\url{https://towardsdatascience.com/inferring-causality-in-time-series-data-b8b75fe52c46}}
  (\bibinfo {year} {2019}),\ \bibinfo {note} {accessed: Aug 28,
  2020}\BibitemShut {NoStop}%
\bibitem [{\citenamefont {Lungarella}\ \emph {et~al.}(2007)\citenamefont
  {Lungarella}, \citenamefont {Ishiguro}, \citenamefont {Kuniyoshi},\ and\
  \citenamefont {Otsu}}]{Lungarella2007-vs}%
  \BibitemOpen
  \bibfield  {author} {\bibinfo {author} {\bibfnamefont {M.}~\bibnamefont
  {Lungarella}}, \bibinfo {author} {\bibfnamefont {K.}~\bibnamefont
  {Ishiguro}}, \bibinfo {author} {\bibfnamefont {Y.}~\bibnamefont {Kuniyoshi}},
  \ and\ \bibinfo {author} {\bibfnamefont {N.}~\bibnamefont {Otsu}},\
  }\bibfield  {title} {\enquote {\bibinfo {title} {Methods for quantifying the
  causal structure of bivariate time series},}\ }\href@noop {} {\bibfield
  {journal} {\bibinfo  {journal} {Int. J. Bifurcat. Chaos}\ }\textbf {\bibinfo
  {volume} {17}},\ \bibinfo {pages} {903--921} (\bibinfo {year}
  {2007})}\BibitemShut {NoStop}%
\bibitem [{\citenamefont {Chen}\ \emph {et~al.}(2004)\citenamefont {Chen},
  \citenamefont {Rangarajan}, \citenamefont {Feng},\ and\ \citenamefont
  {Ding}}]{Chen2004-vu}%
  \BibitemOpen
  \bibfield  {author} {\bibinfo {author} {\bibfnamefont {Y.}~\bibnamefont
  {Chen}}, \bibinfo {author} {\bibfnamefont {G.}~\bibnamefont {Rangarajan}},
  \bibinfo {author} {\bibfnamefont {J.}~\bibnamefont {Feng}}, \ and\ \bibinfo
  {author} {\bibfnamefont {M.}~\bibnamefont {Ding}},\ }\bibfield  {title}
  {\enquote {\bibinfo {title} {Analyzing multiple nonlinear time series with
  extended {Granger} causality},}\ }\href@noop {} {\bibfield  {journal}
  {\bibinfo  {journal} {Phys. Lett. A}\ }\textbf {\bibinfo {volume} {324}},\
  \bibinfo {pages} {26--35} (\bibinfo {year} {2004})}\BibitemShut {NoStop}%
\bibitem [{\citenamefont {Ancona}, \citenamefont {Marinazzo},\ and\
  \citenamefont {Stramaglia}(2004)}]{Ancona2004-cr}%
  \BibitemOpen
  \bibfield  {author} {\bibinfo {author} {\bibfnamefont {N.}~\bibnamefont
  {Ancona}}, \bibinfo {author} {\bibfnamefont {D.}~\bibnamefont {Marinazzo}}, \
  and\ \bibinfo {author} {\bibfnamefont {S.}~\bibnamefont {Stramaglia}},\
  }\bibfield  {title} {\enquote {\bibinfo {title} {Radial basis function
  approach to nonlinear {Granger} causality of time series},}\ }\href@noop {}
  {\bibfield  {journal} {\bibinfo  {journal} {Phys. Rev. E Stat. Nonlin. Soft
  Matter Phys.}\ }\textbf {\bibinfo {volume} {70}},\ \bibinfo {pages} {056221}
  (\bibinfo {year} {2004})}\BibitemShut {NoStop}%
\bibitem [{\citenamefont {Feldmann}\ and\ \citenamefont
  {Bhattacharya}(2004)}]{Feldmann2004-yt}%
  \BibitemOpen
  \bibfield  {author} {\bibinfo {author} {\bibfnamefont {U.}~\bibnamefont
  {Feldmann}}\ and\ \bibinfo {author} {\bibfnamefont {J.}~\bibnamefont
  {Bhattacharya}},\ }\bibfield  {title} {\enquote {\bibinfo {title}
  {Predictability improvement as an asymmetrical measure of interdependence in
  bivariate time series},}\ }\href@noop {} {\bibfield  {journal} {\bibinfo
  {journal} {Int. J. Bifurcat. Chaos}\ }\textbf {\bibinfo {volume} {14}},\
  \bibinfo {pages} {505--514} (\bibinfo {year} {2004})}\BibitemShut {NoStop}%
\bibitem [{\citenamefont {Marschinski}\ and\ \citenamefont
  {Kantz}(2002)}]{Marschinski2002-zh}%
  \BibitemOpen
  \bibfield  {author} {\bibinfo {author} {\bibfnamefont {R.}~\bibnamefont
  {Marschinski}}\ and\ \bibinfo {author} {\bibfnamefont {H.}~\bibnamefont
  {Kantz}},\ }\bibfield  {title} {\enquote {\bibinfo {title} {Analysing the
  information flow between financial time series},}\ }\href@noop {} {\bibfield
  {journal} {\bibinfo  {journal} {The European Physical Journal B - Condensed
  Matter and Complex Systems}\ }\textbf {\bibinfo {volume} {30}},\ \bibinfo
  {pages} {275--281} (\bibinfo {year} {2002})}\BibitemShut {NoStop}%
\bibitem [{\citenamefont {Barnett}, \citenamefont {Barrett},\ and\
  \citenamefont {Seth}(2009)}]{Barnett2009-se}%
  \BibitemOpen
  \bibfield  {author} {\bibinfo {author} {\bibfnamefont {L.}~\bibnamefont
  {Barnett}}, \bibinfo {author} {\bibfnamefont {A.~B.}\ \bibnamefont
  {Barrett}}, \ and\ \bibinfo {author} {\bibfnamefont {A.~K.}\ \bibnamefont
  {Seth}},\ }\bibfield  {title} {\enquote {\bibinfo {title} {Granger causality
  and transfer entropy are equivalent for gaussian variables},}\ }\href@noop {}
  {\bibfield  {journal} {\bibinfo  {journal} {Phys. Rev. Lett.}\ }\textbf
  {\bibinfo {volume} {103}},\ \bibinfo {pages} {238701} (\bibinfo {year}
  {2009})}\BibitemShut {NoStop}%
\bibitem [{\citenamefont {Marinazzo}, \citenamefont {Pellicoro},\ and\
  \citenamefont {Stramaglia}(2008)}]{Marinazzo2008-os}%
  \BibitemOpen
  \bibfield  {author} {\bibinfo {author} {\bibfnamefont {D.}~\bibnamefont
  {Marinazzo}}, \bibinfo {author} {\bibfnamefont {M.}~\bibnamefont
  {Pellicoro}}, \ and\ \bibinfo {author} {\bibfnamefont {S.}~\bibnamefont
  {Stramaglia}},\ }\bibfield  {title} {\enquote {\bibinfo {title} {Kernel
  method for nonlinear granger causality},}\ }\href@noop {} {\bibfield
  {journal} {\bibinfo  {journal} {Phys. Rev. Lett.}\ }\textbf {\bibinfo
  {volume} {100}},\ \bibinfo {pages} {144103} (\bibinfo {year}
  {2008})}\BibitemShut {NoStop}%
\bibitem [{\citenamefont {Palu{\v s}}\ \emph {et~al.}(2001)\citenamefont
  {Palu{\v s}}, \citenamefont {Kom{\'a}rek}, \citenamefont {Hrnc{\'\i}r},\ and\
  \citenamefont {Sterbov{\'a}}}]{Palus2001-po}%
  \BibitemOpen
  \bibfield  {author} {\bibinfo {author} {\bibfnamefont {M.}~\bibnamefont
  {Palu{\v s}}}, \bibinfo {author} {\bibfnamefont {V.}~\bibnamefont
  {Kom{\'a}rek}}, \bibinfo {author} {\bibfnamefont {Z.}~\bibnamefont
  {Hrnc{\'\i}r}}, \ and\ \bibinfo {author} {\bibfnamefont {K.}~\bibnamefont
  {Sterbov{\'a}}},\ }\bibfield  {title} {\enquote {\bibinfo {title}
  {Synchronization as adjustment of information rates: detection from bivariate
  time series},}\ }\href@noop {} {\bibfield  {journal} {\bibinfo  {journal}
  {Phys. Rev. E Stat. Nonlin. Soft Matter Phys.}\ }\textbf {\bibinfo {volume}
  {63}},\ \bibinfo {pages} {046211} (\bibinfo {year} {2001})}\BibitemShut
  {NoStop}%
\bibitem [{\citenamefont {Kraskov}, \citenamefont {St{\"o}gbauer},\ and\
  \citenamefont {Grassberger}(2004)}]{Kraskov2004-sf}%
  \BibitemOpen
  \bibfield  {author} {\bibinfo {author} {\bibfnamefont {A.}~\bibnamefont
  {Kraskov}}, \bibinfo {author} {\bibfnamefont {H.}~\bibnamefont
  {St{\"o}gbauer}}, \ and\ \bibinfo {author} {\bibfnamefont {P.}~\bibnamefont
  {Grassberger}},\ }\href@noop {} {\enquote {\bibinfo {title} {Estimating
  mutual information},}\ } (\bibinfo {year} {2004})\BibitemShut {NoStop}%
\bibitem [{\citenamefont {Takens}(1981)}]{Takens1981-tt}%
  \BibitemOpen
  \bibfield  {author} {\bibinfo {author} {\bibfnamefont {F.}~\bibnamefont
  {Takens}},\ }\bibfield  {title} {\enquote {\bibinfo {title} {Detecting
  strange attractors in turbulence},}\ }in\ \href@noop {} {\emph {\bibinfo
  {booktitle} {Dynamical Systems and Turbulence, Warwick 1980}}}\ (\bibinfo
  {publisher} {Springer Berlin Heidelberg},\ \bibinfo {year} {1981})\ pp.\
  \bibinfo {pages} {366--381}\BibitemShut {NoStop}%
\bibitem [{\citenamefont {Arnhold}\ \emph {et~al.}(1999)\citenamefont
  {Arnhold}, \citenamefont {Grassberger}, \citenamefont {Lehnertz},\ and\
  \citenamefont {Elger}}]{Arnhold1999-si}%
  \BibitemOpen
  \bibfield  {author} {\bibinfo {author} {\bibfnamefont {J.}~\bibnamefont
  {Arnhold}}, \bibinfo {author} {\bibfnamefont {P.}~\bibnamefont
  {Grassberger}}, \bibinfo {author} {\bibfnamefont {K.}~\bibnamefont
  {Lehnertz}}, \ and\ \bibinfo {author} {\bibfnamefont {C.~E.}\ \bibnamefont
  {Elger}},\ }\bibfield  {title} {\enquote {\bibinfo {title} {A robust method
  for detecting interdependences: application to intracranially recorded
  {EEG}},}\ }\href@noop {} {\bibfield  {journal} {\bibinfo  {journal} {Physica
  D}\ }\textbf {\bibinfo {volume} {134}},\ \bibinfo {pages} {419--430}
  (\bibinfo {year} {1999})}\BibitemShut {NoStop}%
\bibitem [{\citenamefont {Bhattacharya}, \citenamefont {Pereda},\ and\
  \citenamefont {Petsche}(2003)}]{Bhattacharya2003-cz}%
  \BibitemOpen
  \bibfield  {author} {\bibinfo {author} {\bibfnamefont {J.}~\bibnamefont
  {Bhattacharya}}, \bibinfo {author} {\bibfnamefont {E.}~\bibnamefont
  {Pereda}}, \ and\ \bibinfo {author} {\bibfnamefont {H.}~\bibnamefont
  {Petsche}},\ }\bibfield  {title} {\enquote {\bibinfo {title} {Effective
  detection of coupling in short and noisy bivariate data},}\ }\href@noop {}
  {\bibfield  {journal} {\bibinfo  {journal} {IEEE Trans. Syst. Man Cybern. B
  Cybern.}\ }\textbf {\bibinfo {volume} {33}},\ \bibinfo {pages} {85--95}
  (\bibinfo {year} {2003})}\BibitemShut {NoStop}%
\bibitem [{\citenamefont {Kennel}, \citenamefont {Brown},\ and\ \citenamefont
  {Abarbanel}(1992)}]{Kennel1992-ej}%
  \BibitemOpen
  \bibfield  {author} {\bibinfo {author} {\bibfnamefont {M.~B.}\ \bibnamefont
  {Kennel}}, \bibinfo {author} {\bibfnamefont {R.}~\bibnamefont {Brown}}, \
  and\ \bibinfo {author} {\bibfnamefont {H.~D.}\ \bibnamefont {Abarbanel}},\
  }\bibfield  {title} {\enquote {\bibinfo {title} {Determining embedding
  dimension for phase-space reconstruction using a geometrical construction},}\
  }\href@noop {} {\bibfield  {journal} {\bibinfo  {journal} {Phys. Rev. A}\
  }\textbf {\bibinfo {volume} {45}},\ \bibinfo {pages} {3403--3411} (\bibinfo
  {year} {1992})}\BibitemShut {NoStop}%
\bibitem [{\citenamefont {Fraser}\ and\ \citenamefont
  {Swinney}(1986)}]{Fraser1986-bc}%
  \BibitemOpen
  \bibfield  {author} {\bibinfo {author} {\bibfnamefont {A.~M.}\ \bibnamefont
  {Fraser}}\ and\ \bibinfo {author} {\bibfnamefont {H.~L.}\ \bibnamefont
  {Swinney}},\ }\bibfield  {title} {\enquote {\bibinfo {title} {Independent
  coordinates for strange attractors from mutual information},}\ }\href@noop {}
  {\bibfield  {journal} {\bibinfo  {journal} {Phys. Rev. A Gen. Phys.}\
  }\textbf {\bibinfo {volume} {33}},\ \bibinfo {pages} {1134--1140} (\bibinfo
  {year} {1986})}\BibitemShut {NoStop}%
\bibitem [{\citenamefont {Tibshirani}, \citenamefont {Walther},\ and\
  \citenamefont {Hastie}(2001)}]{Tibshirani2001-tp}%
  \BibitemOpen
  \bibfield  {author} {\bibinfo {author} {\bibfnamefont {R.}~\bibnamefont
  {Tibshirani}}, \bibinfo {author} {\bibfnamefont {G.}~\bibnamefont {Walther}},
  \ and\ \bibinfo {author} {\bibfnamefont {T.}~\bibnamefont {Hastie}},\
  }\bibfield  {title} {\enquote {\bibinfo {title} {Estimating the number of
  clusters in a data set via the gap statistic},}\ }\href@noop {} {\bibfield
  {journal} {\bibinfo  {journal} {J. R. Stat. Soc. Series B Stat. Methodol.}\
  }\textbf {\bibinfo {volume} {63}},\ \bibinfo {pages} {411--423} (\bibinfo
  {year} {2001})}\BibitemShut {NoStop}%
\bibitem [{\citenamefont {Rissanen}(1978)}]{Rissanen1978-jp}%
  \BibitemOpen
  \bibfield  {author} {\bibinfo {author} {\bibfnamefont {J.}~\bibnamefont
  {Rissanen}},\ }\bibfield  {title} {\enquote {\bibinfo {title} {Modeling by
  shortest data description},}\ }\href@noop {} {\bibfield  {journal} {\bibinfo
  {journal} {Automatica}\ }\textbf {\bibinfo {volume} {14}},\ \bibinfo {pages}
  {465--471} (\bibinfo {year} {1978})}\BibitemShut {NoStop}%
\bibitem [{\citenamefont {Hall}\ and\ \citenamefont
  {Hannan}(1988)}]{Hall1988-mg}%
  \BibitemOpen
  \bibfield  {author} {\bibinfo {author} {\bibfnamefont {P.}~\bibnamefont
  {Hall}}\ and\ \bibinfo {author} {\bibfnamefont {E.~J.}\ \bibnamefont
  {Hannan}},\ }\bibfield  {title} {\enquote {\bibinfo {title} {On stochastic
  complexity and nonparametric density estimation},}\ }\href@noop {} {\bibfield
   {journal} {\bibinfo  {journal} {Biometrika}\ }\textbf {\bibinfo {volume}
  {75}},\ \bibinfo {pages} {705--714} (\bibinfo {year} {1988})}\BibitemShut
  {NoStop}%
\bibitem [{\citenamefont {Clark}\ \emph {et~al.}(2015)\citenamefont {Clark},
  \citenamefont {Ye}, \citenamefont {Isbell}, \citenamefont {Deyle},
  \citenamefont {Cowles}, \citenamefont {Tilman},\ and\ \citenamefont
  {Sugihara}}]{Clark2015-nv}%
  \BibitemOpen
  \bibfield  {author} {\bibinfo {author} {\bibfnamefont {A.~T.}\ \bibnamefont
  {Clark}}, \bibinfo {author} {\bibfnamefont {H.}~\bibnamefont {Ye}}, \bibinfo
  {author} {\bibfnamefont {F.}~\bibnamefont {Isbell}}, \bibinfo {author}
  {\bibfnamefont {E.~R.}\ \bibnamefont {Deyle}}, \bibinfo {author}
  {\bibfnamefont {J.}~\bibnamefont {Cowles}}, \bibinfo {author} {\bibfnamefont
  {G.~D.}\ \bibnamefont {Tilman}}, \ and\ \bibinfo {author} {\bibfnamefont
  {G.}~\bibnamefont {Sugihara}},\ }\href@noop {} {\enquote {\bibinfo {title}
  {Spatial convergent cross mapping to detect causal relationships from short
  time series},}\ } (\bibinfo {year} {2015})\BibitemShut {NoStop}%
\bibitem [{\citenamefont {M{\o}nster}\ \emph {et~al.}(2016)\citenamefont
  {M{\o}nster}, \citenamefont {Fusaroli}, \citenamefont {Tyl{\'e}n},
  \citenamefont {Roepstorff},\ and\ \citenamefont {Sherson}}]{Monster2016-wy}%
  \BibitemOpen
  \bibfield  {author} {\bibinfo {author} {\bibfnamefont {D.}~\bibnamefont
  {M{\o}nster}}, \bibinfo {author} {\bibfnamefont {R.}~\bibnamefont
  {Fusaroli}}, \bibinfo {author} {\bibfnamefont {K.}~\bibnamefont {Tyl{\'e}n}},
  \bibinfo {author} {\bibfnamefont {A.}~\bibnamefont {Roepstorff}}, \ and\
  \bibinfo {author} {\bibfnamefont {J.~F.}\ \bibnamefont {Sherson}},\
  }\bibfield  {title} {\enquote {\bibinfo {title} {Inferring causality from
  noisy time series data},}\ }\href@noop {} {\bibfield  {journal} {\bibinfo
  {journal} {arXiv}\ } (\bibinfo {year} {2016})},\ \bibinfo {note}
  {\url{https://arxiv.org/pdf/1603.01155}},\ \Eprint
  {http://arxiv.org/abs/1603.01155} {arXiv:1603.01155 [nlin.CD]} \BibitemShut
  {NoStop}%
\bibitem [{\citenamefont {H{\'e}non}(1976)}]{Henon1976-kw}%
  \BibitemOpen
  \bibfield  {author} {\bibinfo {author} {\bibfnamefont {M.}~\bibnamefont
  {H{\'e}non}},\ }\bibfield  {title} {\enquote {\bibinfo {title} {A
  two-dimensional mapping with a strange attractor},}\ }\href@noop {}
  {\bibfield  {journal} {\bibinfo  {journal} {Commun. Math. Phys.}\ }\textbf
  {\bibinfo {volume} {50}},\ \bibinfo {pages} {69--77} (\bibinfo {year}
  {1976})}\BibitemShut {NoStop}%
\bibitem [{\citenamefont {Cutts}\ and\ \citenamefont
  {Eglen}(2014)}]{Cutts2014-ow}%
  \BibitemOpen
  \bibfield  {author} {\bibinfo {author} {\bibfnamefont {C.~S.}\ \bibnamefont
  {Cutts}}\ and\ \bibinfo {author} {\bibfnamefont {S.~J.}\ \bibnamefont
  {Eglen}},\ }\bibfield  {title} {\enquote {\bibinfo {title} {Detecting
  pairwise correlations in spike trains: an objective comparison of methods and
  application to the study o retinal waves},}\ }\href@noop {} {\bibfield
  {journal} {\bibinfo  {journal} {J. Neurosci.}\ }\textbf {\bibinfo {volume}
  {34}},\ \bibinfo {pages} {14288--14303} (\bibinfo {year} {2014})}\BibitemShut
  {NoStop}%
\bibitem [{\citenamefont {Siggiridou}\ and\ \citenamefont
  {Kugiumtzis}(2016)}]{Siggiridou2016-yj}%
  \BibitemOpen
  \bibfield  {author} {\bibinfo {author} {\bibfnamefont {E.}~\bibnamefont
  {Siggiridou}}\ and\ \bibinfo {author} {\bibfnamefont {D.}~\bibnamefont
  {Kugiumtzis}},\ }\bibfield  {title} {\enquote {\bibinfo {title} {Granger
  causality in multivariate time series using a {Time-Ordered} restricted
  vector autoregressive model},}\ }\href@noop {} {\bibfield  {journal}
  {\bibinfo  {journal} {IEEE Trans. Signal Process.}\ }\textbf {\bibinfo
  {volume} {64}},\ \bibinfo {pages} {1759--1773} (\bibinfo {year}
  {2016})}\BibitemShut {NoStop}%
\bibitem [{\citenamefont {Guo}\ \emph {et~al.}(2008)\citenamefont {Guo},
  \citenamefont {Seth}, \citenamefont {Kendrick}, \citenamefont {Zhou},\ and\
  \citenamefont {Feng}}]{Guo2008-nx}%
  \BibitemOpen
  \bibfield  {author} {\bibinfo {author} {\bibfnamefont {S.}~\bibnamefont
  {Guo}}, \bibinfo {author} {\bibfnamefont {A.~K.}\ \bibnamefont {Seth}},
  \bibinfo {author} {\bibfnamefont {K.~M.}\ \bibnamefont {Kendrick}}, \bibinfo
  {author} {\bibfnamefont {C.}~\bibnamefont {Zhou}}, \ and\ \bibinfo {author}
  {\bibfnamefont {J.}~\bibnamefont {Feng}},\ }\href@noop {} {\enquote {\bibinfo
  {title} {Partial {Granger} causality---eliminating exogenous inputs and
  latent variables},}\ } (\bibinfo {year} {2008})\BibitemShut {NoStop}%
\bibitem [{\citenamefont {Lizier}(2014)}]{Lizier2014-xt}%
  \BibitemOpen
  \bibfield  {author} {\bibinfo {author} {\bibfnamefont {J.~T.}\ \bibnamefont
  {Lizier}},\ }\bibfield  {title} {\enquote {\bibinfo {title} {{JIDT}: An
  {information-theoretic} toolkit for studying the dynamics of complex
  systems},}\ }\href@noop {} {\bibfield  {journal} {\bibinfo  {journal}
  {Frontiers in Robotics and AI}\ }\textbf {\bibinfo {volume} {1}},\ \bibinfo
  {pages} {11} (\bibinfo {year} {2014})}\BibitemShut {NoStop}%
\bibitem [{\citenamefont {Vlachos}\ and\ \citenamefont
  {Kugiumtzis}(2010)}]{Vlachos2010-ta}%
  \BibitemOpen
  \bibfield  {author} {\bibinfo {author} {\bibfnamefont {I.}~\bibnamefont
  {Vlachos}}\ and\ \bibinfo {author} {\bibfnamefont {D.}~\bibnamefont
  {Kugiumtzis}},\ }\bibfield  {title} {\enquote {\bibinfo {title} {Nonuniform
  state-space reconstruction and coupling detection},}\ }\href@noop {}
  {\bibfield  {journal} {\bibinfo  {journal} {Phys. Rev. E Stat. Nonlin. Soft
  Matter Phys.}\ }\textbf {\bibinfo {volume} {82}},\ \bibinfo {pages} {016207}
  (\bibinfo {year} {2010})}\BibitemShut {NoStop}%
\bibitem [{\citenamefont {Park}\ \emph {et~al.}(2020)\citenamefont {Park},
  \citenamefont {Smith}, \citenamefont {Sugihara},\ and\ \citenamefont
  {Deyle}}]{Park2020-su}%
  \BibitemOpen
  \bibfield  {author} {\bibinfo {author} {\bibfnamefont {J.}~\bibnamefont
  {Park}}, \bibinfo {author} {\bibfnamefont {C.}~\bibnamefont {Smith}},
  \bibinfo {author} {\bibfnamefont {G.}~\bibnamefont {Sugihara}}, \ and\
  \bibinfo {author} {\bibfnamefont {E.}~\bibnamefont {Deyle}},\ }\href@noop {}
  {\enquote {\bibinfo {title} {{EDM}: Empirical dynamic modelling ({'pyEDM'}).
  {Python} package version 1.7.0.}}\ } (\bibinfo {year} {2020}),\ \bibinfo
  {note} {\url{https://github.com/SugiharaLab}}\BibitemShut {NoStop}%
\bibitem [{\citenamefont {Edinburgh}(2021)}]{Edinburgh2021-ha}%
  \BibitemOpen
  \bibfield  {author} {\bibinfo {author} {\bibfnamefont {T.}~\bibnamefont
  {Edinburgh}},\ }\href {\doibase 10.5281/zenodo.4746192} {\enquote {\bibinfo
  {title} {Bivariate causality indices review: code, data and figures},}\
  }\bibinfo {howpublished} {\url{https://doi.org/10.5281/zenodo.4746192}}
  (\bibinfo {year} {2021})\BibitemShut {NoStop}%
\bibitem [{\citenamefont {Brunner}\ \emph {et~al.}(2019)\citenamefont
  {Brunner}, \citenamefont {Ikegwu}, \citenamefont {Trauger},\ and\
  \citenamefont {Trauger}}]{Brunner2019-ip}%
  \BibitemOpen
  \bibfield  {author} {\bibinfo {author} {\bibfnamefont {R.}~\bibnamefont
  {Brunner}}, \bibinfo {author} {\bibfnamefont {K.}~\bibnamefont {Ikegwu}},
  \bibinfo {author} {\bibfnamefont {J.}~\bibnamefont {Trauger}}, \ and\
  \bibinfo {author} {\bibfnamefont {T.}~\bibnamefont {Trauger}},\ }\href@noop
  {} {\enquote {\bibinfo {title} {{PyIF}},}\ }\bibinfo {howpublished}
  {\url{https://github.com/lcdm-uiuc/PyIF}} (\bibinfo {year} {2019}),\ \bibinfo
  {note} {accessed: Sep 05, 2020}\BibitemShut {NoStop}%
\bibitem [{\citenamefont {Nour Jamal El-Din}\ and\ \citenamefont
  {Aljabasini}(2018)}]{NourJamalElDin2020-su}%
  \BibitemOpen
  \bibfield  {author} {\bibinfo {author} {\bibfnamefont {A.}~\bibnamefont {Nour
  Jamal El-Din}}\ and\ \bibinfo {author} {\bibfnamefont {O.}~\bibnamefont
  {Aljabasini}},\ }\href@noop {} {\enquote {\bibinfo {title} {{Kernel Granger
  Causality}},}\ }\bibinfo {howpublished}
  {\url{https://github.com/ITE-5th/fuzzy-clustering}} (\bibinfo {year}
  {2018}),\ \bibinfo {note} {accessed: Jan 15, 2021}\BibitemShut {NoStop}%
\bibitem [{\citenamefont {Perez}\ and\ \citenamefont
  {Granger}(2007)}]{Perez2007-lt}%
  \BibitemOpen
  \bibfield  {author} {\bibinfo {author} {\bibfnamefont {F.}~\bibnamefont
  {Perez}}\ and\ \bibinfo {author} {\bibfnamefont {B.~E.}\ \bibnamefont
  {Granger}},\ }\bibfield  {title} {\enquote {\bibinfo {title} {{IPython}: A
  system for interactive scientific computing},}\ }\href@noop {} {\bibfield
  {journal} {\bibinfo  {journal} {Computing in Science Engineering}\ }\textbf
  {\bibinfo {volume} {9}},\ \bibinfo {pages} {21--29} (\bibinfo {year}
  {2007})}\BibitemShut {NoStop}%
\bibitem [{\citenamefont {Hunter}(2007)}]{Hunter2007-vy}%
  \BibitemOpen
  \bibfield  {author} {\bibinfo {author} {\bibfnamefont {J.~D.}\ \bibnamefont
  {Hunter}},\ }\bibfield  {title} {\enquote {\bibinfo {title} {Matplotlib: A
  {2D} graphics environment},}\ }\href@noop {} {\bibfield  {journal} {\bibinfo
  {journal} {Computing in Science Engineering}\ }\textbf {\bibinfo {volume}
  {9}},\ \bibinfo {pages} {90--95} (\bibinfo {year} {2007})}\BibitemShut
  {NoStop}%
\bibitem [{\citenamefont {Harris}\ \emph {et~al.}(2020)\citenamefont {Harris},
  \citenamefont {Millman}, \citenamefont {van~der Walt}, \citenamefont
  {Gommers}, \citenamefont {Virtanen}, \citenamefont {Cournapeau},
  \citenamefont {Wieser}, \citenamefont {Taylor}, \citenamefont {Berg},
  \citenamefont {Smith}, \citenamefont {Kern}, \citenamefont {Picus},
  \citenamefont {Hoyer}, \citenamefont {van Kerkwijk}, \citenamefont {Brett},
  \citenamefont {Haldane}, \citenamefont {Del~R{\'\i}o}, \citenamefont {Wiebe},
  \citenamefont {Peterson}, \citenamefont {G{\'e}rard-Marchant}, \citenamefont
  {Sheppard}, \citenamefont {Reddy}, \citenamefont {Weckesser}, \citenamefont
  {Abbasi}, \citenamefont {Gohlke},\ and\ \citenamefont
  {Oliphant}}]{Harris2020-od}%
  \BibitemOpen
  \bibfield  {author} {\bibinfo {author} {\bibfnamefont {C.~R.}\ \bibnamefont
  {Harris}}, \bibinfo {author} {\bibfnamefont {K.~J.}\ \bibnamefont {Millman}},
  \bibinfo {author} {\bibfnamefont {S.~J.}\ \bibnamefont {van~der Walt}},
  \bibinfo {author} {\bibfnamefont {R.}~\bibnamefont {Gommers}}, \bibinfo
  {author} {\bibfnamefont {P.}~\bibnamefont {Virtanen}}, \bibinfo {author}
  {\bibfnamefont {D.}~\bibnamefont {Cournapeau}}, \bibinfo {author}
  {\bibfnamefont {E.}~\bibnamefont {Wieser}}, \bibinfo {author} {\bibfnamefont
  {J.}~\bibnamefont {Taylor}}, \bibinfo {author} {\bibfnamefont
  {S.}~\bibnamefont {Berg}}, \bibinfo {author} {\bibfnamefont {N.~J.}\
  \bibnamefont {Smith}}, \bibinfo {author} {\bibfnamefont {R.}~\bibnamefont
  {Kern}}, \bibinfo {author} {\bibfnamefont {M.}~\bibnamefont {Picus}},
  \bibinfo {author} {\bibfnamefont {S.}~\bibnamefont {Hoyer}}, \bibinfo
  {author} {\bibfnamefont {M.~H.}\ \bibnamefont {van Kerkwijk}}, \bibinfo
  {author} {\bibfnamefont {M.}~\bibnamefont {Brett}}, \bibinfo {author}
  {\bibfnamefont {A.}~\bibnamefont {Haldane}}, \bibinfo {author} {\bibfnamefont
  {J.~F.}\ \bibnamefont {Del~R{\'\i}o}}, \bibinfo {author} {\bibfnamefont
  {M.}~\bibnamefont {Wiebe}}, \bibinfo {author} {\bibfnamefont
  {P.}~\bibnamefont {Peterson}}, \bibinfo {author} {\bibfnamefont
  {P.}~\bibnamefont {G{\'e}rard-Marchant}}, \bibinfo {author} {\bibfnamefont
  {K.}~\bibnamefont {Sheppard}}, \bibinfo {author} {\bibfnamefont
  {T.}~\bibnamefont {Reddy}}, \bibinfo {author} {\bibfnamefont
  {W.}~\bibnamefont {Weckesser}}, \bibinfo {author} {\bibfnamefont
  {H.}~\bibnamefont {Abbasi}}, \bibinfo {author} {\bibfnamefont
  {C.}~\bibnamefont {Gohlke}}, \ and\ \bibinfo {author} {\bibfnamefont {T.~E.}\
  \bibnamefont {Oliphant}},\ }\bibfield  {title} {\enquote {\bibinfo {title}
  {Array programming with {NumPy}},}\ }\href@noop {} {\bibfield  {journal}
  {\bibinfo  {journal} {Nature}\ }\textbf {\bibinfo {volume} {585}},\ \bibinfo
  {pages} {357--362} (\bibinfo {year} {2020})}\BibitemShut {NoStop}%
\bibitem [{\citenamefont {Davis}(2012)}]{Davis2012-ip}%
  \BibitemOpen
  \bibfield  {author} {\bibinfo {author} {\bibfnamefont {M.}~\bibnamefont
  {Davis}},\ }\href@noop {} {\enquote {\bibinfo {title} {{Palettable}},}\
  }\bibinfo {howpublished} {\url{https://github.com/jiffyclub/palettable}}
  (\bibinfo {year} {2012}),\ \bibinfo {note} {accessed: Mar 09,
  2020}\BibitemShut {NoStop}%
\bibitem [{\citenamefont {McKinney}(2010)}]{McKinney2010-ey}%
  \BibitemOpen
  \bibfield  {author} {\bibinfo {author} {\bibfnamefont {W.}~\bibnamefont
  {McKinney}},\ }\bibfield  {title} {\enquote {\bibinfo {title} {Data
  structures for statistical computing in python},}\ }in\ \href@noop {} {\emph
  {\bibinfo {booktitle} {Proceedings of the 9th Python in Science
  Conference}}}\ (\bibinfo  {publisher} {SciPy},\ \bibinfo {year}
  {2010})\BibitemShut {NoStop}%
\bibitem [{\citenamefont {Van~Rossum}\ and\ \citenamefont
  {Drake~Jr}(1995)}]{VanRossum1995-aa}%
  \BibitemOpen
  \bibfield  {author} {\bibinfo {author} {\bibfnamefont {G.}~\bibnamefont
  {Van~Rossum}}\ and\ \bibinfo {author} {\bibfnamefont {F.~L.}\ \bibnamefont
  {Drake~Jr}},\ }\href@noop {} {\emph {\bibinfo {title} {Python tutorial}}}\
  (\bibinfo  {publisher} {Centrum voor Wiskunde en Informatica Amsterdam, The
  Netherlands},\ \bibinfo {year} {1995})\BibitemShut {NoStop}%
\bibitem [{\citenamefont {Pedregosa}\ \emph {et~al.}(2011)\citenamefont
  {Pedregosa}, \citenamefont {Varoquaux}, \citenamefont {Gramfort},
  \citenamefont {Michel}, \citenamefont {Thirion}, \citenamefont {Grisel},
  \citenamefont {Blondel}, \citenamefont {Prettenhofer}, \citenamefont {Weiss},
  \citenamefont {Dubourg}, \citenamefont {Vanderplas}, \citenamefont {Passos},
  \citenamefont {Cournapeau}, \citenamefont {Brucher}, \citenamefont {Perrot},\
  and\ \citenamefont {Duchesnay}}]{Pedregosa2011-bt}%
  \BibitemOpen
  \bibfield  {author} {\bibinfo {author} {\bibfnamefont {F.}~\bibnamefont
  {Pedregosa}}, \bibinfo {author} {\bibfnamefont {G.}~\bibnamefont
  {Varoquaux}}, \bibinfo {author} {\bibfnamefont {A.}~\bibnamefont {Gramfort}},
  \bibinfo {author} {\bibfnamefont {V.}~\bibnamefont {Michel}}, \bibinfo
  {author} {\bibfnamefont {B.}~\bibnamefont {Thirion}}, \bibinfo {author}
  {\bibfnamefont {O.}~\bibnamefont {Grisel}}, \bibinfo {author} {\bibfnamefont
  {M.}~\bibnamefont {Blondel}}, \bibinfo {author} {\bibfnamefont
  {P.}~\bibnamefont {Prettenhofer}}, \bibinfo {author} {\bibfnamefont
  {R.}~\bibnamefont {Weiss}}, \bibinfo {author} {\bibfnamefont
  {V.}~\bibnamefont {Dubourg}}, \bibinfo {author} {\bibfnamefont
  {J.}~\bibnamefont {Vanderplas}}, \bibinfo {author} {\bibfnamefont
  {A.}~\bibnamefont {Passos}}, \bibinfo {author} {\bibfnamefont
  {D.}~\bibnamefont {Cournapeau}}, \bibinfo {author} {\bibfnamefont
  {M.}~\bibnamefont {Brucher}}, \bibinfo {author} {\bibfnamefont
  {M.}~\bibnamefont {Perrot}}, \ and\ \bibinfo {author} {\bibfnamefont
  {{\'E}.}~\bibnamefont {Duchesnay}},\ }\bibfield  {title} {\enquote {\bibinfo
  {title} {Scikit-learn: Machine learning in python},}\ }\href@noop {}
  {\bibfield  {journal} {\bibinfo  {journal} {J. Mach. Learn. Res.}\ }\textbf
  {\bibinfo {volume} {12}},\ \bibinfo {pages} {2825--2830} (\bibinfo {year}
  {2011})}\BibitemShut {NoStop}%
\bibitem [{\citenamefont {Virtanen}\ \emph {et~al.}(2020)\citenamefont
  {Virtanen}, \citenamefont {Gommers}, \citenamefont {Oliphant}, \citenamefont
  {Haberland}, \citenamefont {Reddy}, \citenamefont {Cournapeau}, \citenamefont
  {Burovski}, \citenamefont {Peterson}, \citenamefont {Weckesser},
  \citenamefont {Bright}, \citenamefont {van~der Walt}, \citenamefont {Brett},
  \citenamefont {Wilson}, \citenamefont {Millman}, \citenamefont {Mayorov},
  \citenamefont {Nelson}, \citenamefont {Jones}, \citenamefont {Kern},
  \citenamefont {Larson}, \citenamefont {Carey}, \citenamefont {Polat},
  \citenamefont {Feng}, \citenamefont {Moore}, \citenamefont {VanderPlas},
  \citenamefont {Laxalde}, \citenamefont {Perktold}, \citenamefont {Cimrman},
  \citenamefont {Henriksen}, \citenamefont {Quintero}, \citenamefont {Harris},
  \citenamefont {Archibald}, \citenamefont {Ribeiro}, \citenamefont
  {Pedregosa}, \citenamefont {van Mulbregt},\ and\ \citenamefont {{SciPy 1.0
  Contributors}}}]{Virtanen2020-db}%
  \BibitemOpen
  \bibfield  {author} {\bibinfo {author} {\bibfnamefont {P.}~\bibnamefont
  {Virtanen}}, \bibinfo {author} {\bibfnamefont {R.}~\bibnamefont {Gommers}},
  \bibinfo {author} {\bibfnamefont {T.~E.}\ \bibnamefont {Oliphant}}, \bibinfo
  {author} {\bibfnamefont {M.}~\bibnamefont {Haberland}}, \bibinfo {author}
  {\bibfnamefont {T.}~\bibnamefont {Reddy}}, \bibinfo {author} {\bibfnamefont
  {D.}~\bibnamefont {Cournapeau}}, \bibinfo {author} {\bibfnamefont
  {E.}~\bibnamefont {Burovski}}, \bibinfo {author} {\bibfnamefont
  {P.}~\bibnamefont {Peterson}}, \bibinfo {author} {\bibfnamefont
  {W.}~\bibnamefont {Weckesser}}, \bibinfo {author} {\bibfnamefont
  {J.}~\bibnamefont {Bright}}, \bibinfo {author} {\bibfnamefont {S.~J.}\
  \bibnamefont {van~der Walt}}, \bibinfo {author} {\bibfnamefont
  {M.}~\bibnamefont {Brett}}, \bibinfo {author} {\bibfnamefont
  {J.}~\bibnamefont {Wilson}}, \bibinfo {author} {\bibfnamefont {K.~J.}\
  \bibnamefont {Millman}}, \bibinfo {author} {\bibfnamefont {N.}~\bibnamefont
  {Mayorov}}, \bibinfo {author} {\bibfnamefont {A.~R.~J.}\ \bibnamefont
  {Nelson}}, \bibinfo {author} {\bibfnamefont {E.}~\bibnamefont {Jones}},
  \bibinfo {author} {\bibfnamefont {R.}~\bibnamefont {Kern}}, \bibinfo {author}
  {\bibfnamefont {E.}~\bibnamefont {Larson}}, \bibinfo {author} {\bibfnamefont
  {C.~J.}\ \bibnamefont {Carey}}, \bibinfo {author} {\bibfnamefont
  {{\.I}.}~\bibnamefont {Polat}}, \bibinfo {author} {\bibfnamefont
  {Y.}~\bibnamefont {Feng}}, \bibinfo {author} {\bibfnamefont {E.~W.}\
  \bibnamefont {Moore}}, \bibinfo {author} {\bibfnamefont {J.}~\bibnamefont
  {VanderPlas}}, \bibinfo {author} {\bibfnamefont {D.}~\bibnamefont {Laxalde}},
  \bibinfo {author} {\bibfnamefont {J.}~\bibnamefont {Perktold}}, \bibinfo
  {author} {\bibfnamefont {R.}~\bibnamefont {Cimrman}}, \bibinfo {author}
  {\bibfnamefont {I.}~\bibnamefont {Henriksen}}, \bibinfo {author}
  {\bibfnamefont {E.~A.}\ \bibnamefont {Quintero}}, \bibinfo {author}
  {\bibfnamefont {C.~R.}\ \bibnamefont {Harris}}, \bibinfo {author}
  {\bibfnamefont {A.~M.}\ \bibnamefont {Archibald}}, \bibinfo {author}
  {\bibfnamefont {A.~H.}\ \bibnamefont {Ribeiro}}, \bibinfo {author}
  {\bibfnamefont {F.}~\bibnamefont {Pedregosa}}, \bibinfo {author}
  {\bibfnamefont {P.}~\bibnamefont {van Mulbregt}}, \ and\ \bibinfo {author}
  {\bibnamefont {{SciPy 1.0 Contributors}}},\ }\bibfield  {title} {\enquote
  {\bibinfo {title} {{SciPy} 1.0: fundamental algorithms for scientific
  computing in python},}\ }\href@noop {} {\bibfield  {journal} {\bibinfo
  {journal} {Nat. Methods}\ }\textbf {\bibinfo {volume} {17}},\ \bibinfo
  {pages} {261--272} (\bibinfo {year} {2020})}\BibitemShut {NoStop}%
\bibitem [{\citenamefont {Seabold}\ and\ \citenamefont
  {Perktold}(2010)}]{Seabold2010-ab}%
  \BibitemOpen
  \bibfield  {author} {\bibinfo {author} {\bibfnamefont {S.}~\bibnamefont
  {Seabold}}\ and\ \bibinfo {author} {\bibfnamefont {J.}~\bibnamefont
  {Perktold}},\ }\bibfield  {title} {\enquote {\bibinfo {title} {statsmodels:
  Econometric and statistical modeling with python},}\ }in\ \href@noop {}
  {\emph {\bibinfo {booktitle} {9th Python in Science Conference}}}\ (\bibinfo
  {year} {2010})\BibitemShut {NoStop}%
\end{thebibliography}%


\providecommand{\noopsort}[1]{}\providecommand{\singleletter}[1]{#1}%
\begin{thebibliography}{2}%
\makeatletter
\providecommand \@ifxundefined [1]{%
 \@ifx{#1\undefined}
}%
\providecommand \@ifnum [1]{%
 \ifnum #1\expandafter \@firstoftwo
 \else \expandafter \@secondoftwo
 \fi
}%
\providecommand \@ifx [1]{%
 \ifx #1\expandafter \@firstoftwo
 \else \expandafter \@secondoftwo
 \fi
}%
\providecommand \natexlab [1]{#1}%
\providecommand \enquote  [1]{``#1''}%
\providecommand \bibnamefont  [1]{#1}%
\providecommand \bibfnamefont [1]{#1}%
\providecommand \citenamefont [1]{#1}%
\providecommand \href@noop [0]{\@secondoftwo}%
\providecommand \href [0]{\begingroup \@sanitize@url \@href}%
\providecommand \@href[1]{\@@startlink{#1}\@@href}%
\providecommand \@@href[1]{\endgroup#1\@@endlink}%
\providecommand \@sanitize@url [0]{\catcode `\\12\catcode `\$12\catcode
  `\&12\catcode `\#12\catcode `\^12\catcode `\_12\catcode `\%12\relax}%
\providecommand \@@startlink[1]{}%
\providecommand \@@endlink[0]{}%
\providecommand \url  [0]{\begingroup\@sanitize@url \@url }%
\providecommand \@url [1]{\endgroup\@href {#1}{\urlprefix }}%
\providecommand \urlprefix  [0]{URL }%
\providecommand \Eprint [0]{\href }%
\providecommand \doibase [0]{http://dx.doi.org/}%
\providecommand \selectlanguage [0]{\@gobble}%
\providecommand \bibinfo  [0]{\@secondoftwo}%
\providecommand \bibfield  [0]{\@secondoftwo}%
\providecommand \translation [1]{[#1]}%
\providecommand \BibitemOpen [0]{}%
\providecommand \bibitemStop [0]{}%
\providecommand \bibitemNoStop [0]{.\EOS\space}%
\providecommand \EOS [0]{\spacefactor3000\relax}%
\providecommand \BibitemShut  [1]{\csname bibitem#1\endcsname}%
\let\auto@bib@innerbib\@empty
\bibitem [{\citenamefont {Lungarella}\ \emph {et~al.}(2007)\citenamefont
  {Lungarella}, \citenamefont {Ishiguro}, \citenamefont {Kuniyoshi},\ and\
  \citenamefont {Otsu}}]{Lungarella2007-vs}%
  \BibitemOpen
  \bibfield  {author} {\bibinfo {author} {\bibfnamefont {M.}~\bibnamefont
  {Lungarella}}, \bibinfo {author} {\bibfnamefont {K.}~\bibnamefont
  {Ishiguro}}, \bibinfo {author} {\bibfnamefont {Y.}~\bibnamefont {Kuniyoshi}},
  \ and\ \bibinfo {author} {\bibfnamefont {N.}~\bibnamefont {Otsu}},\
  }\bibfield  {title} {\enquote {\bibinfo {title} {Methods for quantifying the
  causal structure of bivariate time series},}\ }\href@noop {} {\bibfield
  {journal} {\bibinfo  {journal} {Int. J. Bifurcat. Chaos}\ }\textbf {\bibinfo
  {volume} {17}},\ \bibinfo {pages} {903--921} (\bibinfo {year}
  {2007})}\BibitemShut {NoStop}%
\bibitem [{\citenamefont {Kaiser}\ and\ \citenamefont
  {Schreiber}(2002)}]{Kaiser2002-hr}%
  \BibitemOpen
  \bibfield  {author} {\bibinfo {author} {\bibfnamefont {A.}~\bibnamefont
  {Kaiser}}\ and\ \bibinfo {author} {\bibfnamefont {T.}~\bibnamefont
  {Schreiber}},\ }\bibfield  {title} {\enquote {\bibinfo {title} {Information
  transfer in continuous processes},}\ }\href@noop {} {\bibfield  {journal}
  {\bibinfo  {journal} {Physica D}\ }\textbf {\bibinfo {volume} {166}},\
  \bibinfo {pages} {43--62} (\bibinfo {year} {2002})}\BibitemShut {NoStop}%
\end{thebibliography}%

\end{document}


\preprint{AIP/123-QED}

\title[Supplementary materials: Causality indices for bivariate time series data]{Supplementary materials for Causality indices for bivariate time series data: a comparative review of performance}

\author{Tom Edinburgh}
\email{te269@cam.ac.uk}
\author{Stephen J. Eglen}
\affiliation{Department of Applied Mathematics and Theoretical Physics, University of Cambridge, Cambridge CB3 0WA, UK}
\author{Ari Ercole}
\affiliation{Division of Anaesthesia, Department of Medicine, University of Cambridge, Cambridge CB2 0QQ, UK}

\maketitle

\renewcommand\thefigure{S\arabic{figure}}
\renewcommand\thetable{S.\Roman{table}}

\section*{Supplementary materials}

\subsection{Tables and figures}

\begin{figure}[b]
  \centering
  \includegraphics[width=8.5cm]{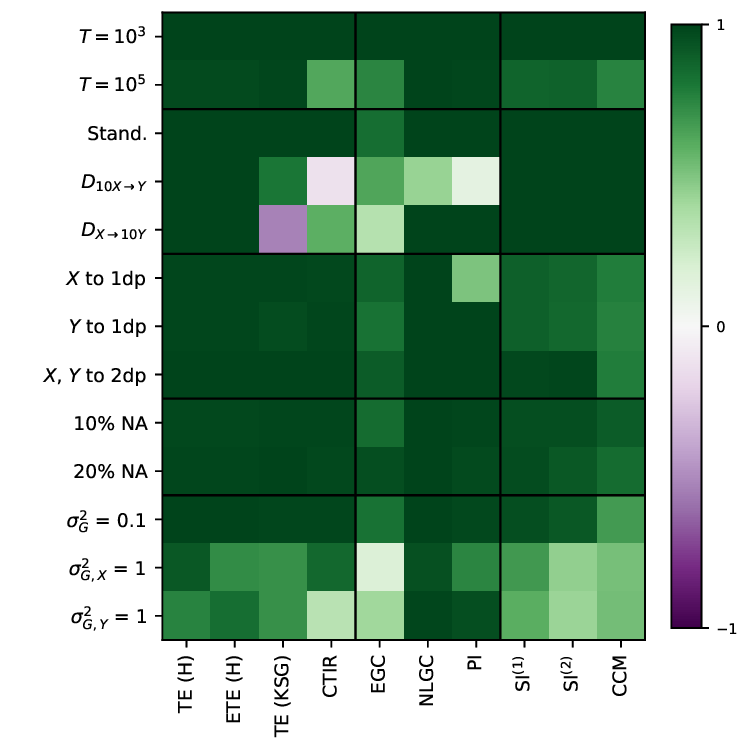}
  \caption{Correlations in $D_{X\rightarrow Y}$ values between the base case of Ulam lattice simulation with $T = 10^3$ and each experiment into the effects of data size, scaling, rounding error, missing data and Gaussian noise, for each method.}
  \label{fig:corr_transformations}
\end{figure}

\begin{figure*}[p]
  \centering
  \includegraphics[width=16cm]{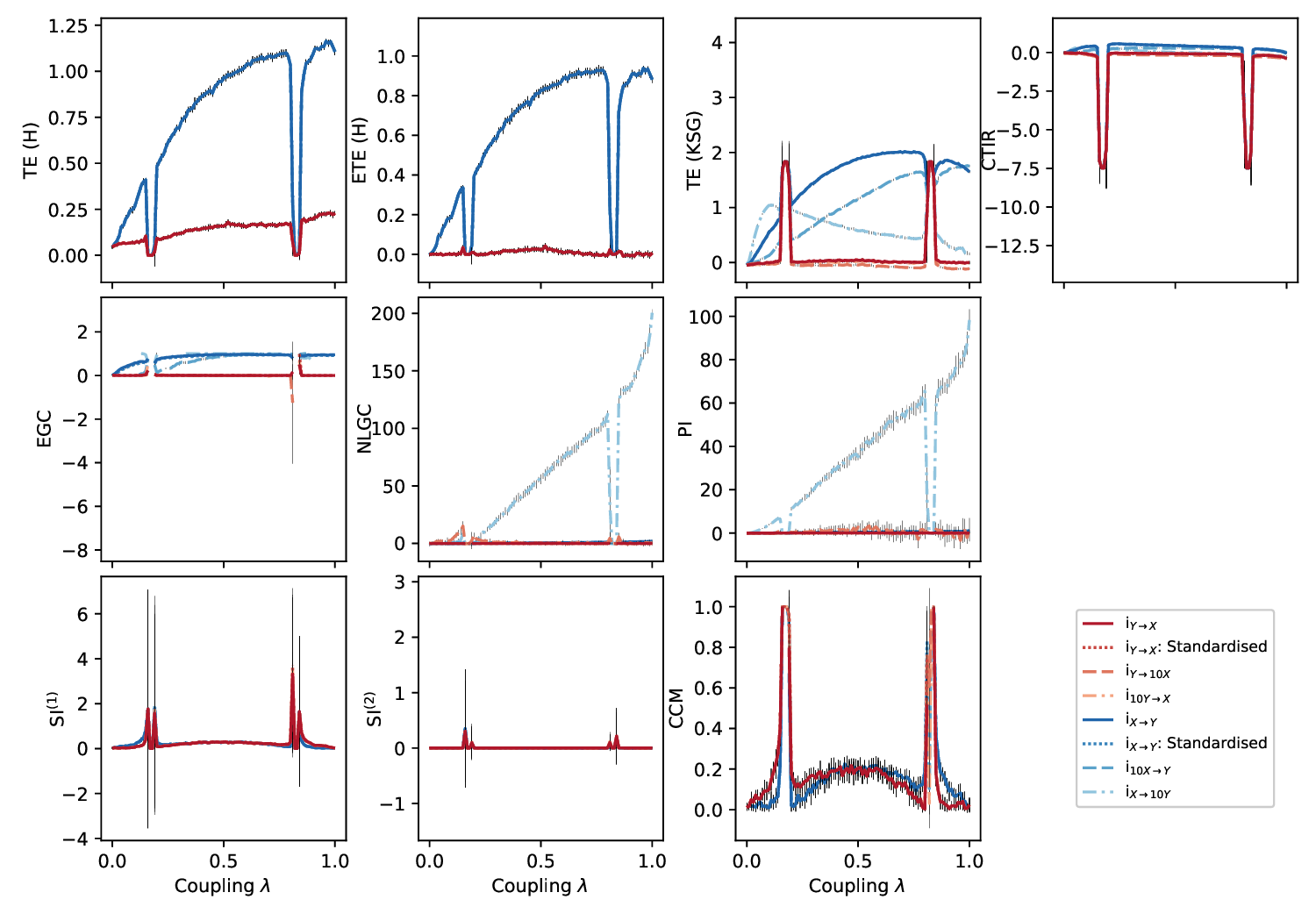}
  \includegraphics[width=16cm]{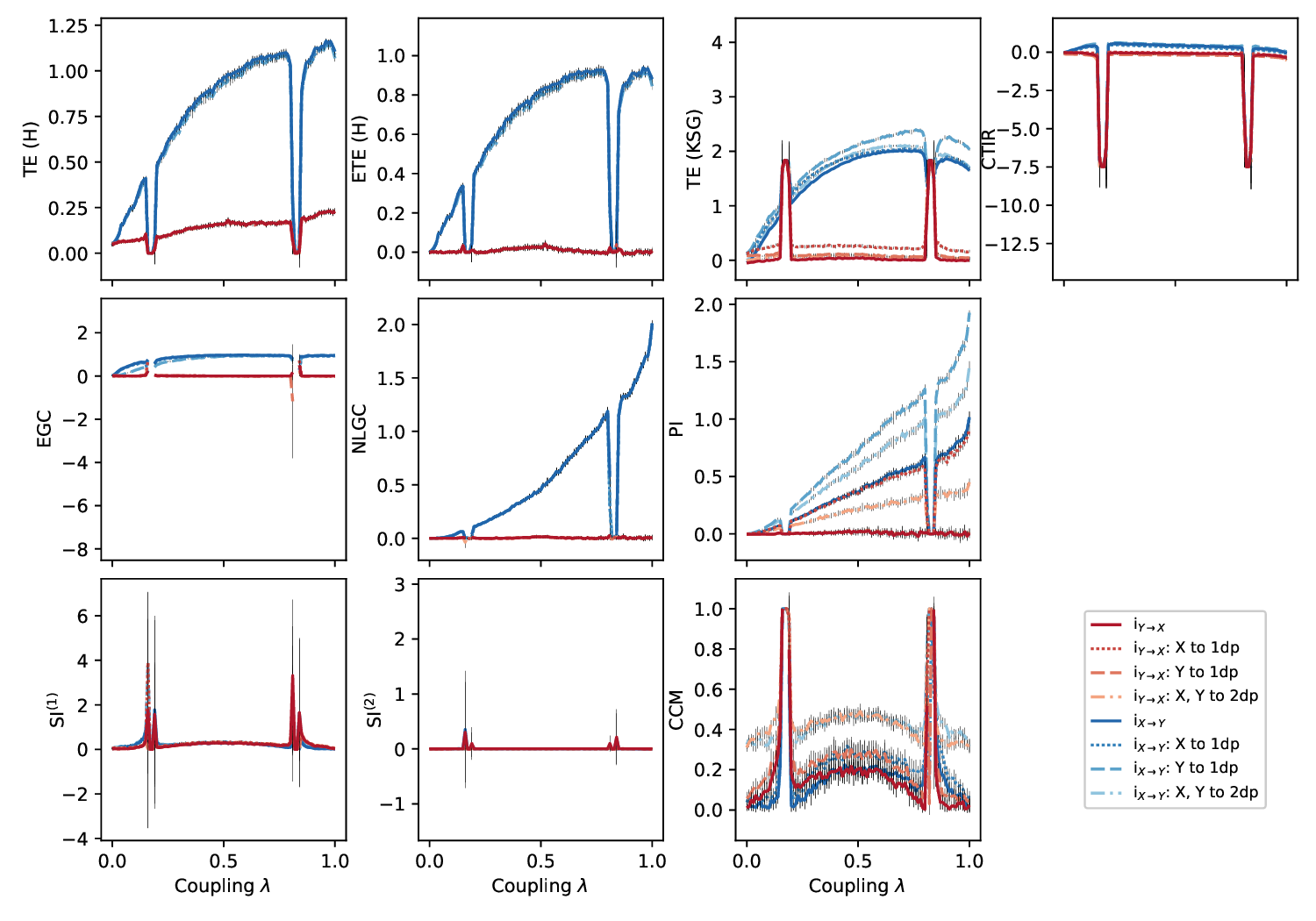}
  \caption{Ulam lattice simulation results with unidirectional ($X\!\rightarrow\! Y$) coupling, showing the effects of scaling and standardisation (top) and rounding error (bottom, $T = 10^3$ data points). Error bars report $\pm$1 standard deviation from mean values, after 10 independent simulations of the Ulam lattice. Simulation parameters are given in Table~2 and parameters for each causality index are given in Table~\ref{tab:indices}. Due to extreme results in the EGC index (for $T = 10^5$ only) when the system exhibits synchrony ($\lambda \approx 0.18, 0.82$), we set these values set to NA.}
  \label{fig:ult1}
\end{figure*}

\begin{figure*}[p]
  \centering
  \includegraphics[width=16cm]{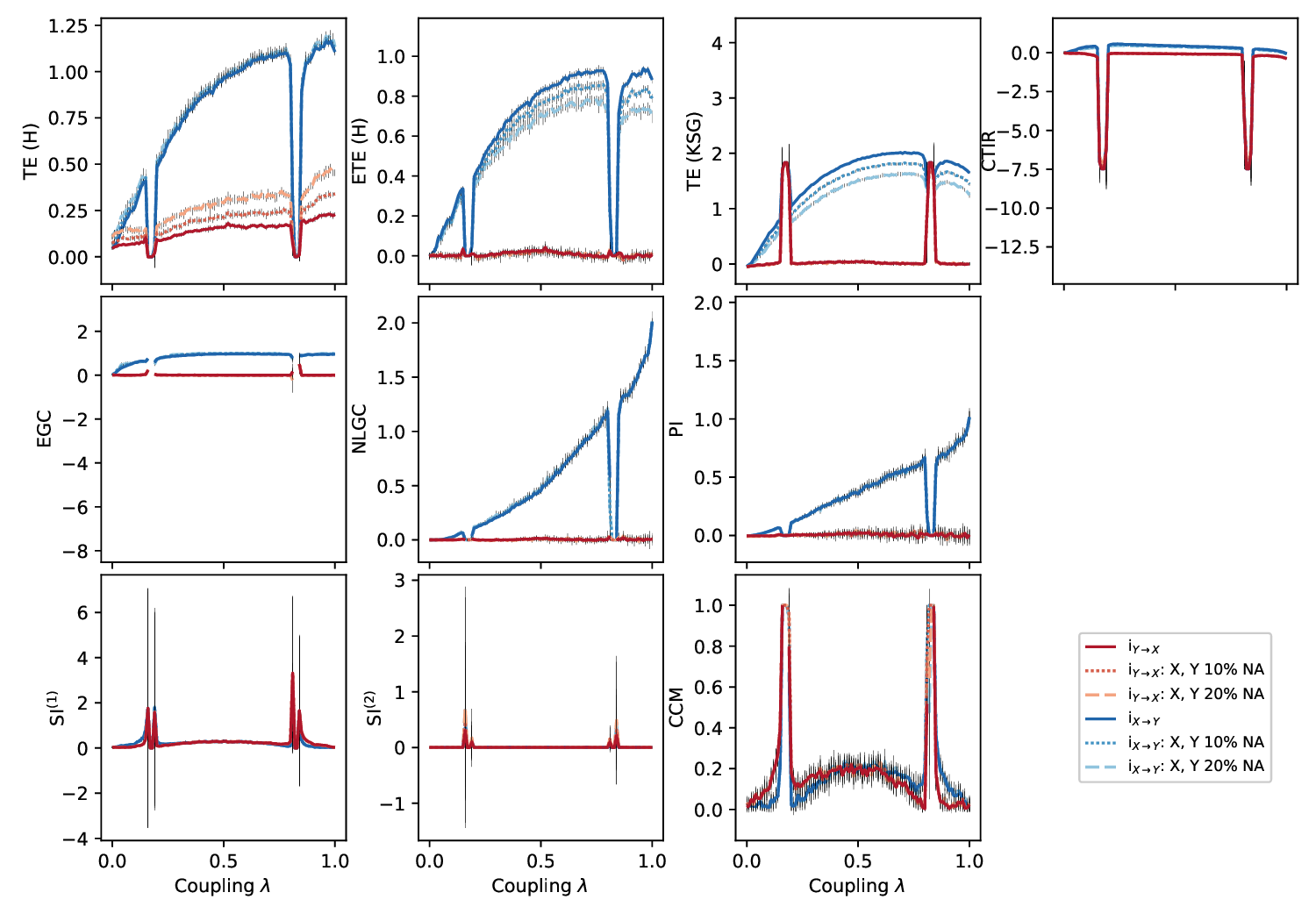}
  \includegraphics[width=16cm]{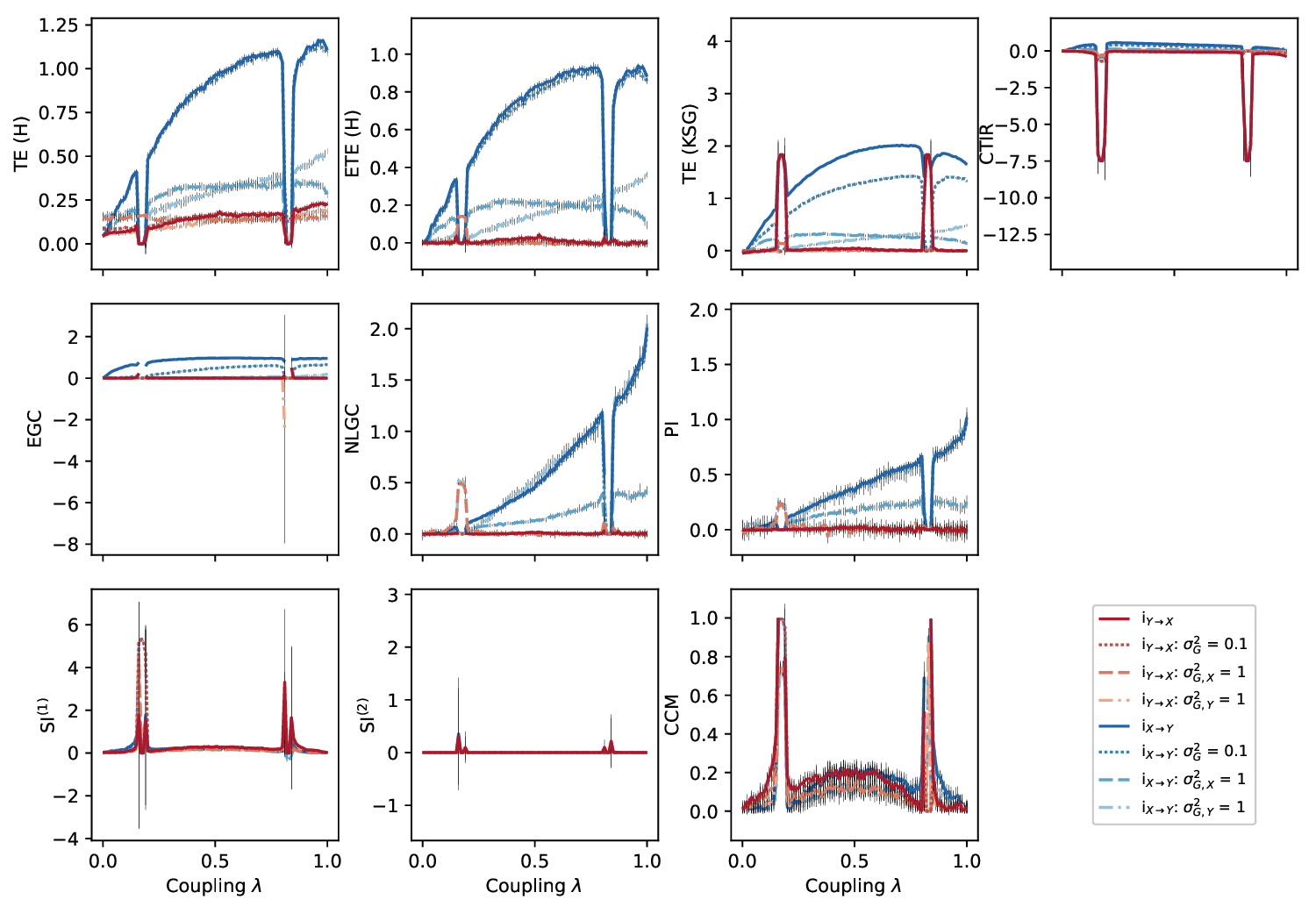}
  \caption{Ulam lattice simulation results with $T = 10^3$ data points and unidirectional ($X\!\rightarrow\! Y$) coupling, showing the effects of missing data (top) and Gaussian noise (bottom). Note $\sigma^2_G$ is the variance of Gaussian noise added to both variables, and $\sigma^2_{G,X}$ is the same added only to $X$ (same for $Y$). Error bars report $\pm$1 standard deviation from mean values, after 10 independent simulations of the Ulam lattice. Simulation parameters are given in Table~2 and parameters for each causality index are given in Table~\ref{tab:indices}.}
  \label{fig:ult2}
\end{figure*}

Table~\ref{tab:indices} shows the causality index parameters for each method and simulated model system. Table~\ref{tab:time} shows the mean and standard deviation of the computational time for each method across all simulated model systems. Figure~\ref{fig:corr_transformations} shows correlations across the additional Ulam lattice experiments for each method. The correlations are between the $D_{X\rightarrow Y}$ values of the `baseline' $T = 10^3$ Ulam lattice experiment and the $D_{X\rightarrow Y}$ values of the experiment on the y-axis, for all $\lambda$. Figures~\ref{fig:ult1}-\ref{fig:ult2} show the results of these experiments in full (except those in the main text).

\begin{table*}[ht]
\caption{Causality index parameter values for each simulation, which are the same as those in Ref~\onlinecite{Lungarella2007-vs}. The indices are as follows (where GC is Granger causality): extended GC (EGC), nonlinear GC (NLGC), predictability improvement (PI), transfer entropy (TE), effective transfer entropy (ETE), coarse-grained transinformation rate (CTIR), similarity indices (SI) and convergent cross mapping (CCM). TE (H) denotes transfer entropy estimation using a histogram partition, and TE (KSG) denotes transfer entropy using Kraskov-St{\"o}gbauer-Grassberger estimation. Common parameters $m$ and $\tau~(*)$ are for all methods, except when otherwise specified in the row corresponding to a given method. CTIR is the only method that does not use these two parameters.}
\begin{tabular}{l|l|l|l|l|l|l|l|l}
    \noalign{\hrule height 2pt}
    Simulation & Linear process & \multicolumn{2}{l|}{Ulam lattice} & \multicolumn{3}{l|}{H{\' e}non unidirectional} & H{\' e}non B (I) & H{\' e}non B (NI) \\
    \cline{2-9}
    $T = 10^p$ & $10^4$ & $10^3$ & $10^5$ & $10^3$ & $10^4$ & $10^5$ & $10^4$ & $10^4$ \\
    \noalign{\hrule height 1.5pt}
    All ($*$) & $m = 2,~\tau = 1$ & \multicolumn{2}{l|}{$m = 1,~\tau = 1$} & \multicolumn{5}{l}{$m = 2,~\tau = 1$} \\
     \cline{1-9}
    EGC & $L = 20$ & \multicolumn{7}{l}{$L = 100$} \\
    & $\delta = 0.8$ & $\delta = 0.5$ & $\delta = 0.2$ & $\delta = 0.5$ & $\delta = 0.3$ & $\delta = 0.2$ & \multicolumn{2}{l}{$\delta = 0.6$} \\
    NLGC & $P = 10$ & \multicolumn{4}{l|}{$P = 50$} & $P = 100$ & \multicolumn{2}{l}{$P = 10$} \\
    & \multicolumn{8}{l}{$\sigma = 0.05$} \\ 
    PI & \multicolumn{8}{l}{$h = 1$} \\
    & $m = 1,~R = 10$ & \multicolumn{7}{l}{$R = 1$} \\
    TE (H) & \multicolumn{8}{l}{$m = 1,~N = 8$} \\ 
    ETE (H) & \multicolumn{8}{l}{$m = 1,~N = 8,~N_{\textnormal{shuffle}} = 10$} \\ 
    TE (KSG) & \multicolumn{8}{l}{$m = 1,~k = 4$} \\
    CTIR & \multicolumn{8}{l}{$k = 4$} \\
    & $\tau_{\max} = 20$ & \multicolumn{7}{l}{$\tau_{\max} = 5$}  \\ 
    SI$^{(2)}$ & $R = 10$ & \multicolumn{7}{l}{$R = 20$} \\
    SI$^{(3)}$ & $R = 30$ & \multicolumn{5}{l|}{$R = 20$} & \multicolumn{2}{l}{$R = 100$} \\
    CCM & \multicolumn{8}{l}{$T_{\max} = T,~n_T = 40,~\delta_\rho = 0.05$} \\
    \noalign{\hrule height 2pt}
\end{tabular}
\label{tab:indices}
\end{table*}

\begin{table*}[ht]
\caption{Computational requirements of each method on each of the datasets, showing the time taken in seconds for the computation of each index and simulation pair, averaged over both the number of runs for all coupling parameter values. Bracketed values give one standard deviation. Methods are as follows (where GC is Granger causality): extended GC (EGC), nonlinear GC (NLGC), predictability improvement (PI), transfer entropy (TE), effective transfer entropy (ETE), coarse-grained transinformation rate (CTIR), similarity indices (SI) and convergent cross mapping (CCM). Simulations as are follows: linear process (LP), Ulam lattice (LP), H{\' e}non unidirectional map (HU), identical H{\' e}non bidirectional map (HB (I)), non-identical H{\' e}non bidirectional map (HB (NI)). TE (H) denotes transfer entropy estimation using a histogram partition, and TE (KSG) denotes transfer entropy using Kraskov-St{\"o}gbauer-Grassberger estimation. ETE (H) includes computation of TE (H) as well. Both SI values are computed concurrently and the computational time listed is for both indices combined.}\vspace{2pt}
\begin{tabular}{l|r|r|r|r|r|r|r|r}
    \noalign{\hrule height 2pt}
    & \multicolumn{1}{c|}{LP} & \multicolumn{2}{c|}{UL} & \multicolumn{3}{c|}{HU} & \multicolumn{1}{c|}{HB (I)} & \multicolumn{1}{c}{HB (NI)} \\
    \cline{2-9}
    \multicolumn{1}{c|}{Method} & \multicolumn{1}{r|}{$T = 10^4$} & \multicolumn{1}{r|}{$T = 10^3$} & \multicolumn{1}{r|}{$T = 10^5$} & \multicolumn{1}{r|}{$T = 10^3$} & \multicolumn{1}{r|}{$T = 10^4$} & \multicolumn{1}{r|}{$T = 10^5$} & \multicolumn{1}{r|}{$T = 10^4$} & $T = 10^4$ \\
    \noalign{\hrule height 2pt}
    EGC & \phantom{0}0.193 (0.008) & 0.131 (0.016) & 4.095 \phantom{0}(12.783) & 0.070 (0.076) & 0.227 (0.010) & 1.216 \phantom{0}(0.452) & 0.250 (0.032) & 0.291 (0.114) \\
    NLGC & 0.693 (0.050) & 0.314 (0.091) & 7.488 \phantom{00}(2.106) & 0.462 (0.025) & 1.536 (0.087) & 27.718 \phantom{0}(4.105) & 0.319 (0.049) & 0.320 (0.124) \\ 
    PI & 0.555 (0.030) & 0.047 (0.006) & 21.323 \phantom{0}(61.590) & 0.051 (0.002) & 0.535 (0.025) & 6.166 \phantom{0}(1.105) & 0.535 (0.056) & 0.675 (0.393) \\ 
    ETE (H) & \textbf{0.041} (\textbf{0.002}) & 0.032 (0.026) & \textbf{1.060} \phantom{00}(\textbf{0.738}) & \textbf{0.013} (\textbf{0.001}) & \textbf{0.039} (\textbf{0.002}) & \textbf{0.370} \phantom{0}(\textbf{0.031}) & \textbf{0.039} (\textbf{0.002}) & \textbf{0.040} (\textbf{0.002}) \\ 
    TE (KSG) & 0.277 (0.012) & \textbf{0.021} (0.005) & 12.461 \phantom{0}(34.498) & 0.018 (\textbf{0.001}) & 0.216 (0.015) & 3.097 \phantom{0}(0.680) & 0.200 (0.019) & 0.264 (0.176) \\
    CTIR & 8.729 (0.394) & 0.181 (0.029) & 140.876 (417.764) & 0.156 (0.011) & 1.860 (0.104) & 25.541 \phantom{0}(2.283) & 1.714 (0.173) & 2.317 (1.624) \\ 
    SI$^{(1,2)}$ & 3.030 (0.129) & 0.086 (0.017) & 108.029 \phantom{0}(22.085) & 0.118 (0.005) & 2.856 (0.120) & 88.200 (50.279) & 3.301 (0.201) & 3.463 (0.199) \\
    CCM & 0.069 (0.003) & 0.028 (\textbf{0.004}) & 4.500 \phantom{0}(15.039) & 0.029 (\textbf{0.001}) & 0.065 (0.003) & 0.740 \phantom{0}(0.171) & 0.068 (0.009) & 0.101 (0.110) \\
    \noalign{\hrule height 2pt}
\end{tabular}
\label{tab:time}
\end{table*}

\subsection{Information theoretic indices with multivariate Gaussian processes}

Transfer entropy has a closed form solution for autoregressive linear process with Gaussian noise, defined ass
\begin{eqnarray}
    x_{t+1} = b_x x_t + \lambda y_t + \epsilon_{x,t},~~~y_{t+1} = b_y y_t + \epsilon_{y,t} \label{eq:sim1} \\
    \epsilon_{x,t} \sim N(0,\sigma^2_x),~~~\epsilon_{y,t} \sim N(0,\sigma^2_y) \nonumber
\end{eqnarray}
With covariance matrices $C_{i,j}(\mathbf{v}) = c(v_i, v_j) = E[v_iv_j] - E[v_i]E[v_j]$ $\mathbf{v} \in \mathbb{R}^m$ (Ref~\onlinecite{Kaiser2002-hr}), the transfer entropy is:
\begin{eqnarray}
     \textnormal{TE}_{Y\rightarrow X} = \frac{1}{2}\log\frac{\det C(\mathbf{x}_t \oplus x_{t+1})\det C(\mathbf{x}_t \oplus \mathbf{y}_t)}{\det C(\mathbf{x}_t \oplus \mathbf{y}_t \oplus x_{t+1})\det C(\mathbf{x}_t)}
\end{eqnarray}
We show algebraic calculations for $m = 1,~\tau = 1$. We set $u = (1 - b_y^2) / \sigma^2_y,~v = 1 - b_x^2, w = 1 - b_x b_y$ (note, these are slightly different to the $u,v,w$ in Ref~\onlinecite{Kaiser2002-hr}). Each Gaussian variable $\epsilon_{x,t}$ and $\epsilon_{y,t}$ is independent of all others. The elements of the covariance matrices are as follows (equations (3)-(8)):
\begin{eqnarray}
    c(y_t, y_t) &=& c(y_{t+1}, y_{t+1}) = c(b_y y_t + \epsilon_{y,t},~b_y y_t + \epsilon_{y,t}) \nonumber \\
    &=& b_y^2 c(y_t, y_t) + c(\epsilon_{y,t},\epsilon_{y,t}) = \frac{\sigma_y^2}{1 - b_y^2} = \frac{1}{u} \\
    c(y_t, y_{t+1}) &=& c(y_t, b_y y_t + \epsilon_{y,t}) = b_y c(y_t, y_t) \nonumber \\
    &=& \frac{b_y\sigma_y^2}{1 - b_y^2} = \frac{b_y}{u} \\
    c(x_t, y_t) &=& c(x_{t+1}, y_{t+1}) = c(b_x x_t + \lambda y_t + \epsilon_{x,t},~b_y y_t + \epsilon_{y,t}) \nonumber \\
    &=& b_x b_y c(x_t, y_t) + \lambda b_y c(y_t, y_t) \nonumber \\
    &=& \frac{1}{1 - b_x b_y} \lambda b_y \frac{\sigma^2_y}{1 - b_y^2} = \frac{\lambda b_y}{uw} \\
    c(x_{t+1},y_t) &=& c(b_x x_t + \lambda y_t + \epsilon_{x,t},~y_t) = b_x c(x_t, y_t) + \lambda c(y_t, y_t) \nonumber \\
    &=& b_x \frac{\lambda b_y}{1 - b_x b_y} \frac{\sigma^2_y}{1 - b_y^2} + \lambda \frac{\sigma_y^2}{1 - b_y^2} = \frac{\lambda}{uw} \nonumber \\
    c(x_t, y_{t+1}) &=& c(x_t, b_y y_t + \epsilon_{y, t}) = b_y c(x_t, y_t) \nonumber \\
    &=& b_y \frac{\lambda b_y}{1 - b_x b_y} \frac{\sigma^2_y}{1 - b_y^2} = \frac{\lambda b_y^2}{uw}
\end{eqnarray}
Continuing with covariance elements:
\begin{eqnarray}
    c(x_t, x_t) &=& c(x_{t+1}, x_{t+1}) \nonumber \\
    &=& c(b_x x_t + \lambda y_t + \epsilon_{x,t}, b_x x_t + \lambda y_t + \epsilon_{x,t}) \nonumber \\
    &=& b_x^2 c(x_t, x_t) + 2\lambda b_x c(x_t, y_t) + \lambda^2 c(y_t, y_t) \nonumber \\
    &=& \frac{1}{1 - b_x^2}\left[ 2\lambda b_x \frac{\lambda b_y}{1 - b_x b_y} \frac{\sigma^2_y}{1 - b_y^2} + \lambda^2 \frac{\sigma_y^2}{1 - b_y^2} + \sigma^2_x \right] \nonumber \\
    &=& \frac{1}{1 - b_x^2}\left(\sigma^2_x + \frac{\lambda^2 (1 + b_xb_y) \sigma^2_y}{(1 - b_xb_y)(1 - b_y^2)}\right) \nonumber \\
    &=& \frac{1}{uvw}(uw \sigma^2_x + \lambda^2 (1 + b_x b_y)) \\
    c(x_t, x_{t+1}) &=& c(x_t, b_x x_t + \lambda y_t + \epsilon_{x,t}) = b_x c(x_t, x_t) + \lambda c(x_t, y_t) \nonumber
\end{eqnarray}
\begin{eqnarray}
    &=& \frac{b_x}{1 - b_x^2}\left(\sigma^2_x + \frac{\lambda^2 (1 + b_xb_y) \sigma^2_y}{(1 - b_xb_y)(1 - b_y^2)}\right) \nonumber \\
    &&~~+ \frac{\lambda^2 b_y}{1 - b_x b_y} \frac{\sigma^2_y}{1 - b_y^2} \nonumber \\
    &=& \frac{1}{1 - b_x^2}\left(b_x \sigma^2_x + \frac{\lambda^2 (b_x + b_y) \sigma^2_y}{(1 - b_xb_y)(1 - b_y^2)}\right) \nonumber \\
    &=& \frac{1}{uvw}(b_x uw \sigma^2_x + \lambda^2 (b_x + b_y))
\end{eqnarray}
As an exercise, we compute covariance matrices of the different subspaces involved:
\begin{widetext}
\begin{eqnarray}
    \det C(x_t \oplus x_{t + 1}) &=& c(x_t, x_t) c(x_{t+1}, x_{t+1}) - c(x_t, x_{t+1})^2 \nonumber \\
    &=& \frac{1}{(uvw)^2}\left[u^2 w^2 \sigma^4_x (1 - b_x^2) + 2\lambda^2 \sigma^2_x uw (1 + b_xb_y - b_xb_y - b_x^2) + \lambda^4 (1 + 2b_xb_y + b_x^2 b_y^2 - b_x^2 - 2b_xb_y - b_y^2)\right] \nonumber \\
    &=& \frac{1}{u v w^2}\left[ u w^2 \sigma^4_x + 2\lambda^2 w \sigma^2_x + \lambda^4 \sigma^2_y \right] \\
    \det C(y_t \oplus y_{t + 1}) &=& c(y_t, y_t) c(y_{t + 1}, y_{t + 1}) - c(y_t, y_{t + 1})^2 = \frac{1}{u^2} (1 - b_y^2) = \frac{\sigma^2_y}{u} = \sigma^2_y~C(y_t) \\
    \det C(x_t \oplus y_t) &=& c(x_t, x_t) c(y_t, y_t) - c(x_t, y_t)^2 = \frac{1}{u^2 v w^2}\left[ uw^2 \sigma^2_x + \lambda^2 w - \lambda^2 b_x b_y w - \lambda^2 b_y^2 v \right] \nonumber \\
    &=& \frac{1}{u^2 v w^2}\left[ uw^2 \sigma^2_x + \lambda^2 ((1 - b_x b_y)(1 - b_x b_y) - b_y^2 (1 - b_x^2)) \right] = \frac{1}{uvw^2}\left( w^2 \sigma^2_x + \lambda^2 \sigma^2_y \right) \\[0.5em]
    \det C(x_t \oplus y_t \oplus x_{t + 1}) &=& c(x_t, x_t) c(y_t, y_t) c(x_{t+1}, x_{t+1}) + 2 c(x_t, x_{t+1}) c(x_{t+1}, y_t) c(x_t, y_t) \nonumber \\
    &&~~~~~~- c(x_t, x_{t+1})^2 c(y_t, y_t) - c(x_t, x_t) c(x_{t+1}, y_t)^2 - c(x_{t+1}, x_{t+1})c(x_t, y_t)^2 \nonumber \\[0.5em]
    &=& \frac{1}{u^3 v^2 w^3} \left[ w(uw \sigma^2_x + \lambda^2(1 + b_xb_y))^2 + 2\lambda^2 b_y v(b_x uw \sigma^2_x + \lambda^2(b_x + b_y)) \right. \nonumber \\
    &&~~~~\left. - \lambda^2 b_y^2 v(uw \sigma^2_x + \lambda^2(1 + b_x b_y)) - w(b_x uw \sigma^2_x + \lambda^2(b_x + b_y))^2 - \lambda^2 v (uw \sigma^2_x + \lambda^2 (1 + b_x b_y))\right] \nonumber \\[0.5em]
    &=& \frac{1}{u^3 v^2 w^3} \left[ u^2 w^3 (1 - b_x^2) \sigma^4_x + \lambda^2 uw \sigma^2_x \Big(2w(1 + b_xb_y) + 2b_x b_y v - b_y^2 v - 2b_x w(b_x + b_y) - v \Big) \right. \nonumber \\
    &&~~~~\left. + \lambda^4 \Big(w(1 + b_x b_y)^2 + 2b_y v(b_x + b_y) - b_y^2 v(1 + b_x b_y) - w (b_x + b_y)^2 - v(1 + b_x b_y) \Big) \right] \nonumber \\[0.5em]
    &=& \frac{1}{u^3 v^2 w^3} \left[ u^2 v w^3 \sigma^4_x + \lambda^2 uw \sigma^2_x \Big(2w(2 - w) + 2(1 - w) v - (1 - u \sigma^2_y) v - 2 w(1 - v + 1 - w) - v \Big) \right. \nonumber \\
    &~&~~~~\left. + \lambda^4 \Big(w(1 + 2b_x b_y + b_x^2 b_y^2 - b_x^2 - 2b_x b_y - b_y^2) + v(2b_x b_y + 2b_y^2 - b_y^2 - b_x b_y^3 - 1 - b_xb_y) \Big) \right] \nonumber \\[0.5em]
    &\stackrel{(\ddagger)}{=}& \frac{1}{u^3 v^2 w^3}\left[ u^2 v w^3 \sigma^4_x + \lambda^2 u^2 v w \sigma^2_x \sigma^2_y \right] = \frac{\sigma^2_x}{u v w^2}\left( w^2 \sigma^2_x + \lambda^2 \sigma^2_y \right) = \sigma^2_x~\det C(x_t \oplus y_t) \\[0.5em]
    (\ddagger)~~0 &=& \lambda^4 \Big(w (1 - b_x^2) (1 - b_y^2) + v(- 1 + b_x b_y)(1 - b_y^2)\Big) = \lambda^4 (uvw \sigma^2_y - uvw \sigma^2_y) \\[0.5em]
    \det C(x_t \oplus y_t \oplus y_{t + 1})  &=& c(x_t, x_t) c(y_t, y_t) c(y_{t+1}, y_{t+1}) + 2 c(x_t, y_t) c(y_t, y_{t+1}) c(x_t, y_{t+1}) \nonumber \\
    &&~~~~~~- c(x_t, y_{t+1})^2 c(y_t, y_t) - c(x_t, x_t) c(y_t, y_{t+1})^2 - c(y_{t+1}, y_{t+1})c(x_t, y_t)^2 \nonumber \\[0.5em]
    &=& \frac{1}{u^3 w^2} \left[ uw^2 c(x_t, x_t) + 2 \lambda^2 b_y^4 - \lambda^2 b_y^4 - b_y^2 uw^2 c(x_t, x_t) - \lambda^2 b_y^2 \right] \nonumber \\
    &=& \frac{1}{u^3w^2} (1 - b_y^2)[uw^2 c(x_t, x_t) - \lambda^2 b_y^2] = \frac{\sigma^2_y}{u} c(x_t, x_t) - \sigma^2_y \left(\frac{\lambda b_y}{uw}\right)^2 \nonumber \\
    &=& \sigma^2_y (c(x_t, x_t) c(y_t, y_t) - c(x_t, y_t)^2) = \sigma^2_y~\det C(x_t \oplus y_t)
\end{eqnarray}
\end{widetext}
In these calculations, we implicitly assume stationarity. For example, if $y_0 = \epsilon_{y} \sim N(0, \sigma^2_0)$, then in reality we have:
\begin{eqnarray*}
    c(y_t, y_t) = \sum_{k = 0}^{t-1} (b_y^{2k} c(\epsilon_{k,t}, \epsilon_{k,t})) + b_y^{2t} \sigma^2_0 = \frac{\sigma^2_y}{1 - b_y^2} + R_{y,t} \\
    R_{y,t} = \sum_{k = t}^{\infty} b_y^{2k} c(\epsilon_{k,t}, \epsilon_{k,t}) - b_y^{2t} \sigma^2_0,~~~\forall t~\geq~1
\end{eqnarray*}After discarding e.g. $10^5$ initial transients, the effect of the initialised states and error $R_{y,t}$ is negligible for all $b_y < 1$. 

Then:
\begin{equation}
    \textnormal{TE}_{X\rightarrow Y} = (1 / 2) \log (1) = 0
\end{equation}
We also have:
\begin{eqnarray}
    &&\textnormal{TE}_{Y\rightarrow X} = \frac{1}{2}\log\frac{\det C(x_t \oplus x_{t+1})\det C(x_t \oplus y_t)}{\det C(x_t \oplus y_t \oplus x_{t+1})\det C(x_t)} \\
    &&= \frac{1}{2}\log\frac{\det C(x_t \oplus x_{t+1})}{\sigma^2_x\det C(x_t)} \nonumber \\
    &&= \frac{1}{2}\log\frac{uw^2 \sigma^4_x + 2\lambda^2 w \sigma^2_x + \lambda^4 \sigma^2_y}{uw^2 \sigma^4_x + \lambda^2 w(2 - w) \sigma^2_x} \nonumber \\
    &&= \frac{1}{2}\log\frac{(1 - b_y^2)(1 - b_x b_y)^2 \sigma^4_x + 2\lambda^2 (1 - b_x b_y) \sigma^2_x \sigma^2_y + \lambda^4 \sigma^4_y}{(1 - b_y^2)(1 - b_x b_y)^2 \sigma^4_x + \lambda^2 (1 - b_x^2 b_y^2) \sigma^2_x \sigma^2_y} \nonumber \\
\end{eqnarray}
\begin{eqnarray}    
    &&\textnormal{TE}_{Y\rightarrow X} = \frac{1}{2}\frac{2(1 - b_x b_y)\sigma^2_x \sigma^2_y - (1 - b_x^2 b_y^2)\sigma^2_x \sigma^2_y}{(1 - b_y^2)(1 - b_x b_y)^2 \sigma^4_x} \lambda^2 + O(\lambda^3) \nonumber\\
    &&= \frac{\sigma^2_y}{\sigma^2_x}\frac{1}{2(1 - b_y^2)} \lambda^2 + O(\lambda^3)
\end{eqnarray}
These computations can be extended to find (conditional) mutual information for arbitrary lags (integer-valued) $\tau$ (and checked using induction). Using the same methods as above, it can be shown that:
\begin{eqnarray}
    c(y_t, y_{t + \tau}) &=& b_y c(y_t, y_{t + \tau - 1}) + c(y_t, \epsilon_{y,t + \tau - 1}) = \frac{\sigma_y^2 b_y^\tau}{1 - b_y^2} \\
    c(x_t, y_{t + \tau}) &=& b_y c(x_t, y_{t + \tau - 1}) + c(x_t, \epsilon_{y,t + \tau - 1}) = b_y^\tau c(x_y, y_t) \nonumber \\
    &=& \frac{\sigma_y^2 b_y^{\tau + 1} \lambda}{(1 - b_y^2)(1 - b_xb_y)} \\
    c(x_t, x_{t + \tau}) &=& b_x c(x_t, x_{t + \tau - 1}) + \lambda c(x_t, y_{t + \tau - 1}) + c(x_t, \epsilon_{x,t + \tau - 1}) \nonumber \\
    &=& \frac{\sigma_x^2 b_x^\tau}{1 - b_x^2} + \frac{\lambda^2 \sigma_y^2}{(1 - b_x^2)(1 - b_y^2)(1 - b_xb_y)} \times A \nonumber \\[0.5em]
    A &=& \left[ (1 - b_x^2) \sum_{k = 0}^\tau b_y^k b_x^{\tau - k} + b_x^{\tau + 1} (b_x + b_y) \right]
\end{eqnarray}
\begin{eqnarray}
    c(y_t, x_{t + \tau}) &=& b_x c(y_t, x_{t + \tau - 1}) + \lambda c(y_t, y_{t + \tau - 1}) + c(y_t, \epsilon_{x, t + \tau - 1}) \nonumber\\
    &=& \frac{\lambda \sigma_y^2}{(1 - b_y^2)(1 - b_x b_y)} \times B
    \nonumber\\
    B &=& \left[ b_x^\tau b_y + (1 - b_xb_y) \sum_{k = 0}^{\tau - 1} b_y^k b_x^{\tau - k}\right] 
\end{eqnarray}CTIR is a much more complicated expression than transfer entropy so we do not provide a closed form solution in full. However, we can form covariance matrices which have elements that are one of the above four expressions, and numerically compute the covariance matrix and their determinants in each case. Finally, a sum over $\tau$ allows us to compute the theoretical value of CTIR for multivariate Gaussian processes.

\bibliography{references}